\documentclass[useAMS,usenatbib]{mn2e}

\usepackage{longtable}
\usepackage{graphics,graphicx}
\usepackage{lscape}

\title[Formamide]{Shedding light on the formation of the pre-biotic molecule formamide with ASAI}
\author[L\'opez-Sepulcre et al.]{A. L\'opez-Sepulcre$^{1,2,3}$\thanks{E-mail:
ana@taurus.phys.s.u-tokyo.ac.jp},  Ali A. Jaber$^{1,2,4}$, E. Mendoza$^{5}$, B. Lefloch$^{1,2}$, C. Ceccarelli$^{1,2}$,
\newauthor
C. Vastel$^{6,7}$, R. Bachiller$^{8}$, J. Cernicharo$^{9}$, C. Codella$^{10}$, C. Kahane$^{1,2}$, M. Kama$^{11}$, 
\newauthor
M. Tafalla$^{8}$
\thanks{Based on observations carried out with the IRAM~30-m Telescope. IRAM is supported by INSU/CNRS (France), MPG (Germany) and IGN (Spain).}\\
\thanks{Based on analysis carried out with the CASSIS software.}\\
$^{1}$Univ. Grenoble Alpes, IPAG, F-38000 Grenoble, France\\
$^{2}$CNRS, IPAG, F-38000 Grenoble, France\\
$^{3}$Department of Physics, The University of Tokyo, Bunkyo-ku, Tokyo 113-0033, Japan\\
$^{4}$University of AL-Muthana, AL-Muthana, Iraq\\
$^{5}$Observat\'orio do Valongo, Universidade Federal do Rio de Janeiro - UFRJ, Rio de Janeiro, Brazil\\
$^{6}$Universit\'e de Toulouse, UPS-OMP, IRAP, Toulouse, France\\
$^{7}$CNRS, IRAP, 9 Av. colonel Roche, BP 44346, 31028 Toulouse Cedex 4, France\\
$^{8}$IGN Observatorio Astron\'omico Nacional (IGN), Calle Alfonso XII, 3. 28014 Madrid, Spain\\
$^{9}$LAM, CAB-CSIC/INTA, Ctra de Torrej\'on a Ajalvir km 4, 28850 Torrej\'on de Ardoz, Madrid, Spain\\
$^{10}$INAF, Osservatorio Astrofisico di Arcetri, Largo E. Fermi 5, 50125, Firenze, Italy\\
$^{11}$Leiden Observatory, P.O. Box 9513, NL-2300 RA, Leiden, The Netherlands}
\begin{document}

\date{Accepted. Received.}

\pagerange{\pageref{firstpage}--\pageref{lastpage}} \pubyear{}

\maketitle

\label{firstpage}

\begin{abstract}
Formamide (NH$_2$CHO) has been proposed as a pre-biotic precursor with a key role in the emergence of life on Earth. While this molecule has been observed in space, most of its detections correspond to high-mass star-forming regions. Motivated by this lack of investigation in the low-mass regime, we searched for formamide, as well as isocyanic acid (HNCO), in 10 low- and intermediate-mass pre-stellar and protostellar objects. The present work is part of the IRAM Large Programme ASAI (Astrochemical Surveys At IRAM), which makes use of unbiased broadband spectral surveys at millimetre wavelengths. We detected HNCO in all the sources and NH$_2$CHO in five of them. We derived their abundances and analysed them together with those reported in the literature for high-mass sources. For those sources with formamide detection, we found a tight and almost linear correlation between HNCO and NH$_2$CHO abundances, with their ratio being roughly constant --between 3 and 10-- across 6 orders of magnitude in luminosity. This suggests the two species are chemically related. The sources without formamide detection, which are also the coldest and devoid of hot corinos, fall well off the correlation, displaying a much larger amount of HNCO relative to NH$_2$CHO. Our results suggest that, while HNCO can be formed in the gas phase during the cold stages of star formation, NH$_2$CHO forms most efficiently on the mantles of dust grains at these temperatures, where it remains frozen until the temperature rises enough to sublimate the icy grain mantles. We propose hydrogenation of HNCO as a likely formation route leading to NH$_2$CHO.
\end{abstract}


\section[intro]{Introduction}

One of the major questions regarding the origin of life on Earth is whether the original chemical mechanism that led from simple molecules to life was connected to metabolism or to genetics, both intimately linked in living beings. Formamide (NH$_2$CHO) contains the four most important elements for biological systems: C, H, O, and N, and it has recently been proposed as a pre-biotic precursor of both metabolic and genetic material, suggesting a common chemical origin for the two mechanisms (\citealp{sal12}).

Formamide was detected for the first time in space by \cite{rubin71} towards Sgr B2 and later in Orion KL. However, dedicated studies of NH$_2$CHO in molecular clouds have started only very recently, as its potential as a key prebiotic molecule has become more evident. These studies present observations of formamide in a number of massive hot molecular cores (\citealp{bis07}, \citealp{adande11}), the low-mass protostellar object IRAS 16293--2422 (\citealp{kahane13}), and the outflow shock regions L1157-B1 and B2 (\citealp{edgar14}, \citealp{yama12}). Its detection in comet Hale-Bopp has also been reported (\citealp{bm00}). Formamide is therefore present in a variety of star-forming environments, as well as on a comet of the Solar System. Whether this implies an exogenous delivery onto a young Earth in the past is a suggestive possibility that needs more evidence to be claimed.

Establishing the formation route(s) of formamide in space remains a challenge. Different chemical pathways have been proposed, both in the gas-phase (e.g. \citealp{redondo14}) and on grain surfaces (e.g. \citealp{raunier04}, \citealp{jones11}). The present work represents an effort to try to understand the dominant mechanisms that lead to the formation of formamide in the interstellar medium. In particular, it seeks to investigate the possible chemical connection between NH$_2$CHO and HNCO, which was proposed by \cite{edgar14}. To this aim, we have performed a homogeneous search of NH$_2$CHO and HNCO in a representative sample of 10 star-forming regions (SFRs) of low- to intermediate-mass type, since most of the formamide detections so far reported concentrate on high-mass SFRs. This is the first systematic study conducted within the context of the IRAM Large Program ASAI (Astrochemical Surveys At IRAM;  P.I.s: B. Lefloch, R. Bachiller), which is dedicated to millimetre astrochemical surveys of low-mass SFRs with the IRAM~30-m telescope.

The source sample and the observations are described in Sects.~\ref{sample} and \ref{obs}, respectively. Section~\ref{results} presents the spectra and describes the analysis carried out to obtain the abundances of NH$_2$CHO and HNCO. Sect.~\ref{disc} compares the derived abundances with those found in the literature for other SFRs, and discusses the formation mechanisms that are favoured by our results. Our conclusions are summarised in Sect.~\ref{conclusions}.

\begin{table*}
 \centering
 \begin{minipage}{160mm}
  \caption{Source sample and their properties.}
  \begin{tabular}{@{}lccccccll@{}}
  \hline
   Source & R.A.(J2000) & Dec.(J2000) & $V_\mathrm{lsr}$ & $d$ & $M$ & $L_\mathrm{bol}$ & Type$^\ast$ & References\\
    &  &  & (km\,s$^{-1}$) & (pc) & (M$_\odot$) & (L$_\odot$) &  & \\
 \hline
 \multicolumn{9}{c}{ASAI}\\
 \hline
L1544 & 05:04:17.21 & +25:10:42.8 & +7.3 & 140 & 2.7 & 1.0 & PSC & 1,2,3\\
TMC1 & 04:41:41.90 & +25:41:27.1 & +6.0 & 140 & 21 & --- & PSC & 1,4\\
B1 & 03:33:20.80 & +31:07:34.0 & +6.5 & 200 & 1.9 & 1.9 & Class 0 & 5,6\\
L1527 & 04:39:53.89 & +26:03:11.0 & +5.9 & 140 & 0.9 & 1.9 & Class 0, WCCC & 1,7,8\\
L1157-mm & 20:39:06.30 & +68:02:15.8 & +2.6 & 325 & 1.5 & 4.7 & Class 0 & 7,8\\
IRAS~4A & 03:29:10.42 & +31:13:32.2 & +7.2 & 235 & 5.6 & 9.1 & Class 0, HC & 7,8\\
SVS13A & 03:29:03.73 & +31:16:03.8 & +6.0 & 235 & 0.34 & 21 & Class 0/1 & 9,10\\
OMC-2~FIR~4 & 05:35:26.97 & --05:09:54.5 & +11.4 & 420 & 30 & 100 & IM proto-cluster & 11,12\\
Cep~E & 23:03:12.80 & +61:42:26.0 & --10.9 & 730 & 35 & 100 & IM protostar & 13\\
\hline
 \multicolumn{9}{c}{TIMASSS}\\
\hline
I16293 & 16:32:22.6 & --24:28:33 & +4.0 & 120 & 3 & 22 & Class 0, HC & 14,15\\
\hline
\end{tabular}
\\
$^\ast$PSC: pre-stellar core; HC: hot corino; WCCC: warm carbon-chain chemistry; IM: intermediate-mass.\\
References: $^1$\cite{e78}, $^2$\cite{evans01}, $^3$\cite{shirley00}, $^4$\cite{toth04}, $^5$\cite{hirano99}, $^6$\cite{marcelino05}, $^{7}$\cite{kris12}, $^{8}$\cite{karska13}, $^{9}$\cite{hirota08}, $^{10}$\cite{chen09}, $^{11}$\cite{crim09}, $^{12}$\cite{furlan14}, $^{13}$\cite{crim10a}, $^{14}$\cite{loinard08}, $^{15}$\cite{correia04}.
\\
\label{tsample}
\end{minipage}
\end{table*}

\section[sample]{Source sample}\label{sample}

Our source sample consists of 10 well-known pre-stellar and protostellar objects representing different masses and evolutionary states, thus providing a complete view of the various types of objects encountered along the first phases of star formation. Their basic properties are listed in Table~\ref{tsample}. All of them belong to the ASAI source sample except one: the Class~0 protobinary IRAS~16293--2422 (hereafter I16293), whose millimetre spectral survey, TIMASSS (The IRAS16293-2422 Millimeter And Submillimeter Spectral Survey), was published by \cite{caux11}. A dedicated study of Complex Organic Molecules (COMs) in this source, including NH$_2$CHO, was recently carried out by \cite{ali14}.

\section[obs]{Observations and data reduction}\label{obs}

The data presented in this work were acquired with the IRAM~30-m telescope near Pico Veleta (Spain) and consist of unbiased spectral surveys at millimetre wavelengths. These are part of the Large Programme ASAI, whose observations and data reduction procedures will be presented in detail in an article by  Lefloch \& Bachiller (2015, \textit{in prep.}). Briefly, we gathered the spectral data in several observing runs between 2011 and 2014 using the EMIR receivers at 3~mm (80--116~GHz), 2~mm (129--173~GHz), and 1.3~mm (200--276~GHz). The main beam sizes for each molecular line analysed are listed in Tables~\ref{tnh2cho} and \ref{thnco}. The three bands were covered for most of the sources. For Cep~E, additional observations were carried out at 0.9-mm (E330 receiver), while just a few frequencies were covered at 2~mm. The Fourier Transform Spectrometer (FTS) units were connected to the receivers, providing a spectral resolution of 195~kHz, except in the case of L1544, for which we used the FTS50 spectrometer, with a resolution of 50~kHz, to resolve the narrow lines ($\Delta V \sim 0.5$~km~s$^{-1}$) that characterise this region. The observations were performed in wobbler switching mode with a throw of 180$''$.

The data were reduced with the package CLASS90 of the GILDAS software collection.\footnote{http://www.iram.fr/IRAMFR/GILDAS/} Through comparison of line intensities among different scans and between horizontal and vertical polarisations, the calibration uncertainties are estimated to be lower than 10 \% at 3~mm and 20\% in the higher frequency bands. After subtraction of the continuum emission via first-order polynomial fitting, a final spectrum was obtained for each source and frequency band after stitching the spectra from each scan and frequency setting. The intensity was converted from antenna temperature ($T_\mathrm{ant}^\ast$) to main beam temperature ($T_\mathrm{mb}$) using the beam efficiencies provided at the IRAM web site\footnote{http://www.iram.es/IRAMES/mainWiki/Iram30mEfficiencies}. In order to improve the signal-to-noise (S/N) ratio, the 2- and 1-mm ASAI data were smoothed to 0.5~km~s$^{-1}$, except in the case of L1544, for which we kept the original spectral resolution.

For I16293, we used the TIMASSS spectral data obtained with the IRAM~30-m telescope at 1, 2, and 3~mm. A detailed description of the observations and an overview of the dataset are reported in \cite{caux11}.

\section{Results}\label{results}

\subsection{Line spectra}\label{lid}

We searched for formamide (NH$_2$CHO) and isocyanic acid (HNCO) in our dataset using the CASSIS software\footnote{CASSIS has been developed by IRAP-UPS/CNRS (http://cassis.irap.omp.eu)} and the Cologne Database for Molecular Spectroscopy (CDMS\footnote{http://www.astro.uni-koeln.de/cdms/}; \citealp{muller01}, 2005) to identify the lines. For NH$_2$CHO, we detected transitions with upper level energies, $E_\mathrm{up}$, below 150~K, and spontaneous emission coefficients, $A_{ij}$, above $10^{-5}$~s$^{-1}$ and $5 \times 10^{-5}$~s$^{-1}$, respectively for the 2/3-mm and the 1-mm data. For HNCO, we detected transitions with $E_\mathrm{u} < 150$~K and $A_{ij} > 10^{-5}$~s$^{-1}$. Tables~\ref{tnh2cho} and \ref{thnco} list all the NH$_2$CHO and HNCO transitions fulfilling these criteria in the observed millimetre bands, as well as the $3\sigma$ detections for each source. The sources where no NH$_2$CHO lines were detected (see below) are not included in Table~\ref{tnh2cho}. For some sources with not many clear formamide detections (e.g. IRAS4A, Cep~E), we included a few additional lines with peak intensities between 2$\sigma$ to 3$\sigma$, as indicated in the tables. We then fitted the lines with a Gaussian function, and excluded from further analysis those falling well below or above the systemic velocity, and/or displaying too narrow or too broad line widths with respect to the typical values encountered for each source.

Table~\ref{tdet} lists, for each source, the number of NH$_2$CHO and HNCO lines detected and used in our analysis (Sect~\ref{rd}). While HNCO is easily detected in all the sources, NH$_2$CHO remains undetected in five objects: L1544, TMC-1, B1, L1527, and L1157mm. Moreover, in those sources where it is detected, the lines are typically weak (S/N$ \sim 3-5$). OMC-2~FIR~4 has the highest number of detected formamide lines, which are also the most intense. The results from the Gaussian fitting to the detected lines are presented in Tables~\ref{tl1544det} -- \ref{tomc2det}. A sample of lines for all the ASAI sources are shown in Figs.~\ref{fspt1} -- \ref{fspt3}.

\begin{table}
 \centering
  \caption{Number of NH$_2$CHO and HNCO detected lines}
  \begin{tabular}{@{}lcccc@{}}
  \hline
   & \multicolumn{2}{c}{NH$_2$CHO} & \multicolumn{2}{c}{HNCO} \\
  Source & \# & $E_\mathrm{u}$ (K) & \# & $E_\mathrm{u}$ (K) \\
 \hline
L1544$^\mathrm{a}$ & 0 & --- & 2 & 10--16\\
TMC1 & 0 & --- & 3 & 10--16\\
B1 & 0 & --- & 4 & 10--30\\
L1527 & 0 & --- & 4 & 10--30\\
L1157-mm & 0 & --- & 4 & 10--30\\
IRAS~4A & 7 & 15--70 &10 & 10--130\\
SVS13A & 13 & 15--130 & 19 & 10--130\\
OMC-2~FIR~4 & 21 & 10--130 & 9 & 10--100\\
Cep~E & 5 & 10--22 & 5 & 10--85\\
\hline
I16293 & 12 & 10--160 & 16 & 10--95\\
\hline
\end{tabular}
\\
$^\mathrm{a}$Only 3-mm data available.
\label{tdet}
\end{table}

\subsection{Derivation of physical properties}\label{rd}

\subsubsection{Rotational diagram analysis}\label{rot}

In order to determine the excitation conditions --i.e. excitation temperature, column density and, eventually, abundance with respect to H$_2$-- of NH$_2$CHO and HNCO for each source in a uniform way, we employed the CASSIS software to build rotational diagrams. This approach assumes (i) that the lines are optically thin, and (ii) Local Thermodynamic Equilibrium (LTE), meaning that a single Boltzmann temperature, known as \textit{rotational temperature}, describes the relative distribution of the population of all the energy levels for a given molecule. Under these assumptions, the upper-level column density

\begin{equation}
N_u = \frac{8\pi k \nu^2}{h c^3 A_{ul}} \frac{1}{\eta_\mathrm{bf}} \int T_\mathrm{mb} dV
\label{enup}
\end{equation}
and the rotational temperature, $T_\mathrm{rot}$, are related as follows:

\begin{equation}
\ln \frac{N_u}{g_u} = ln N_\mathrm{tot} - ln Q(T_\mathrm{rot}) - \frac{E_u}{k T_\mathrm{rot}}
\label{erot}
\end{equation}
where $k$, $\nu$, $h$, and $c$ are, respectively, Boltzmann's constant, the frequency of the transition, Planck's constant, and the speed of light; $g_\mathrm{u}$ is the degeneracy of the upper level, and $N_\mathrm{tot}$ is the total column density of the molecule. The second fraction in Eq.~\ref{enup} is the inverse of the beam-filling factor. We estimated it assuming sources with a gaussian intensity distribution:

\begin{equation}
\eta_\mathrm{bf} = \frac{\theta_\mathrm{s}^2}{\theta_\mathrm{s}^2+\theta_\mathrm{b}^2}
\label{ebf}
\end{equation}
with $\theta_\mathrm{s}$ and $\theta_\mathrm{b}$ being, respectively, the source and telescope beam sizes. We adopted the source sizes indicated in Table~\ref{trd}. In those sources where a hot ($T > 100$~K) inner region is believed to exist, we considered two possible solutions: (i) the emission originates from a compact size representing this inner region or \textit{hot corino}, which typically shows enhanced abundances of Complex Organic Molecules (COMs); and (ii) the emission homogeneously arises from the entire extended molecular envelope of the protostar. We determined the sizes of the compact hot corino regions either from published interferometric maps (SVS13A) or from the gas density structure, $n(r)$, reported in the literature (I16293, IRAS~4A, OMC-2, Cep~E), as indicated in Table~\ref{trd}. In the latter case, we assumed a size equal to the diameter within which the dust temperature is above 100~K.

\begin{figure}
\centering
  \includegraphics[scale=1]{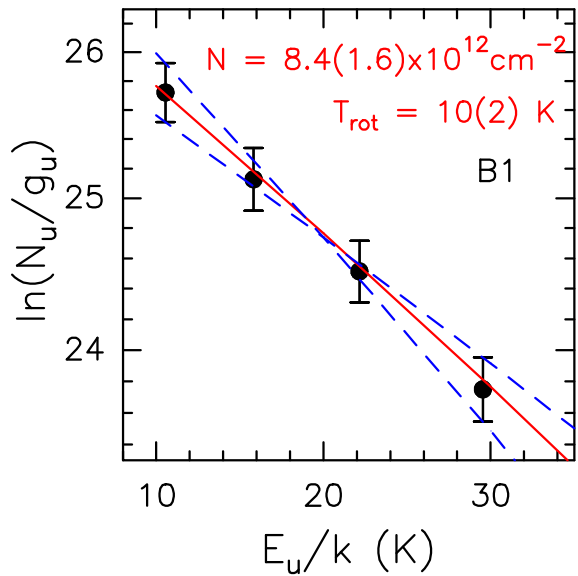} 
  \caption{HNCO rotational diagram of B1. Data points are depicted in black. The red lines correspond to the best fit to the data points. The extreme solutions taking into account the error bars are displayed in dashed blue.}
  \label{frotb1}
\end{figure}

Some sources, such as IRAS~4A, OMC-2, and Cep~E, show extended velocity wings in a few of their lines. In order to separate their contribution to the line emission, we determined their line flux, $\int T_\mathrm{mb} dV$, by fitting a Gaussian function to the affected lines after masking their high-velocity wings. In sources with two to four well-aligned data points in the rotational diagrams, we took into account the relatively large error bars by fitting two additional ``extreme" lines passing through the tips of the error bars of the lowest and largest energy points. An example is shown for B1 in Fig.~\ref{frotb1}, where the two extreme solutions are depicted in blue, while the best solution is marked in red. The remaining rotational diagrams and the best fit to their data points using Eq.~\ref{erot} are shown in Figs.~\ref{frot1} and \ref{frot2}, where the error bars take into account calibration errors as well as the rms value around each line. 

We compared our rotational diagram results with those reported in \cite{marcelino09} for the four sources common to both studies: L1544, TMC-1, B1, and L1527. The column densities of HNCO are in perfect agreement within the uncertainties, while the rotational temperatures agree within 1~K.

For homogeneity with the methodology used for NH$_2$CHO, we estimated the properties of HNCO in the LTE approximation. In addition, by adopting the same source sizes for HNCO and NH$_2$CHO, we assumed that the emission from both molecules originates in the same region(s). The similar average line widths between the two species suggest this is a reasonable assumption. Table~\ref{trd} and Figs.~\ref{frot1} and \ref{frot2} present the results of the rotational diagram analysis. For most of the sources, a single component fits well both the NH$_2$CHO and HNCO points and therefore LTE seems to reproduce well the observations. This can also be seen in Figs.~\ref{fspt1} to \ref{fspt3}, where the observed spectra (in black) and the best fit models (in red) match fairly well. However, for SVS13A, Cep~E, and OMC-2, the compact solutions correspond to HNCO lines that are moderately optically thick ($\tau \sim 1-10$). The most extreme case is Cep~E, for which also the NH$_2$CHO lines are optically thick. This is in contradiction with the underlying assumption of optically thin lines in the rotational diagram method. We find, however, that this caveat can be easily overcome by adopting a slightly larger source size, of 3$''$, 2$''$, and 2$''$, respectively for SVS13A, Cep~E, and OMC-2. Doing this, the resulting column densities are reduced by a factor 2 (OMC-2) to 15 (Cep~E), $\tau$ becomes much smaller than 1, and the lines can be well fitted by the solutions. Consequently, the uncertainties in the compact-solution column densities in these three sources are larger than reported in Table~\ref{trd}, but they are taken into account in the discussion (Sect.~\ref{disc}: see Figs.~\ref{fabu} and \ref{flum}).

There are two objects where a single component does not appear to explain the emission of all the lines: IRAS~4A and I16293, two well-known hot corino sources. Indeed, their rotational diagrams suggest either the contribution of two components, or non-LTE effects, coming into play. Considering the former, Table~\ref{trd} presents the results of a two-component solution to the rotational diagrams of these two objects, where C1 is assumed to represent the cold extended envelope of the protostar, and C2 the small inner hot corino. While this 2-component solution reproduces well the observations, non-LTE effects cannot be ruled out.

\begin{table*}
 \centering
\begin{minipage}{175mm}
  \caption{Results from the rotational diagram analysis of NH$_2$CHO and HNCO: Adopted size and H$_2$ column densities ($N_\mathrm{H2}$), derived rotational temperatures, $T_\mathrm{rot}$, derived HNCO and HN$_2$CHO column densities ($N_\mathrm{HNCO}$, $N_\mathrm{NH2CHO}$), resulting abundances with respect to H$_2$ ( $X_\mathrm{HNCO}$, $X_\mathrm{NH2CHO}$), and ratio of HNCO to NH$_2$CHO column densities ($R$).}
  \begin{tabular}{@{}lccccccccc@{}}
  \hline
   Source & Size$^\mathrm{a}$ & $N_\mathrm{H2}^\mathrm{b}$ & $T_\mathrm{rot(HNCO)}$ & $N_\mathrm{HNCO}$ & $X_\mathrm{HNCO}$ & $T_\mathrm{rot(NH2CHO)}^\mathrm{c}$ & $N_\mathrm{NH2CHO}^\mathrm{c}$ & $X_\mathrm{NH2CHO}$ & $R$\\
    & ($''$) & (10$^{22}$cm$^{-2}$) & (K) & ($10^{12}$cm$^{-2}$) & ($10^{-11}$) & (K) & ($10^{12}$cm$^{-2}$) & ($10^{-11}$) & \\
 \hline
\multicolumn{10}{c}{1-component fit}\\
\hline
L1544 & BF & $9.4 \pm 1.6^1$ & $7 \pm 3$ & $5 \pm 3$ & $5 \pm 3$ & 7 & $< 0.036$ & $< 0.046$ & $> 130$ \\
TMC1$^\mathrm{d}$ & BF & $1.0 \pm 0.1^2$ &  $4 \pm 1$ & $8 \pm 5$ & $80 \pm 50$ & 4 & $< 0.47$ & $< 5.2$ & $> 17$\\
B1 & BF & $7.9 \pm 0.3^3$ & $10 \pm 2$ & $8.4 \pm 1.6$ & $11 \pm 2$ & 10 & $< 0.087$ & $< 0.11$ & $> 97$\\
L1527 & BF & 4.1$^4$ &  $7.5 \pm 1.4$ & $2.5 \pm 1.5$ & $6 \pm 4$ & 7.5 & $< 0.062$ & $< 0.15$ & $> 40$\\
L1157-mm & 30 & 120$^5$ & $8 \pm 1$ & $4 \pm 1$ & $0.35 \pm 0.03$ & 8 & $< 1$ & $< 0.008$ & $> 40$\\
SVS13A (ext)$^\mathrm{e}$ & 20 & 10$^{6}$ & $58 \pm 6$ & $11 \pm 2$ & $11 \pm 2$ & $64 \pm 6$ & $3.0 \pm 0.4$ & $3.0 \pm 0.4$ & $4 \pm 1$\\
\, \, \, \, \, \, \, \, (com)$^\mathrm{e}$ & 1 & 1000$^{7}$ & $36 \pm 3$ & $1500 \pm 300$ & $15 \pm 3$ & $40 \pm 4$ & $320 \pm 60$ & $3.2 \pm 0.6$ & $5 \pm 1$ \\
OMC-2 (ext) & 25 & 19$^8$ & $25 \pm 3$ & $16 \pm 3$ & $1.9 \pm 0.4$ & $58 \pm 4$ & $3.1 \pm 0.2$ & $0.36 \pm 0.02$ & $5 \pm 1$\\
\, \, \, \, \, \, \, (com) & 2 & 4.6$^8$ & $19 \pm 1$ & $900 \pm 100$ & $910 \pm 80$ & $32 \pm 2$ & $110 \pm 10$ & $110 \pm 10$ & $8 \pm 1$ \\
Cep~E (ext) & 40 & 4.8$^9$ & $30 \pm 5$ & $6.2 \pm 0.3$ & $13 \pm 1$ & $9 \pm 2$ & $0.2 \pm 0.1$ & $0.4 \pm 0.2$ & $30 \pm 13$ \\
\, \, \, \, \, \, (com) & 0.5 & 230$^9$ & $17 \pm 1$ & $6000 \pm 1000$ & $130 \pm 15$ & $6 \pm 1$ & $500 \pm 300$ & $11 \pm 5$ & $12 \pm 6$ \\
\hline
\multicolumn{10}{c}{2-component fit}\\
\hline
IRAS~4A (C1) & 30 & 2.9$^{10}$ & $11 \pm 3$ & $10 \pm 1$ & $34 \pm 2$ & $19 \pm 15$ & $0.6 \pm 0.5$ & $1.9 \pm 0.2$ & $18 \pm 2$ \\
IRAS~4A (C2) & 0.5 & 250$^{10}$ & $43 \pm 8$ & $2000 \pm 1000$ & $80 \pm 40$ & $30 \pm 5$ & $500 \pm 100$ & $20 \pm 5$ & $4 \pm 2$ \\
I16293 (C1) & 30 & 2.9$^{11}$ & $14 \pm 5$ & $20 \pm 2$ & $69 \pm 7$ & $5 \pm 1$ & $1.7 \pm 0.6$ & $6 \pm 2$ & $12 \pm 4$ \\
I16293 (C2) & 1.2 & 53$^{11}$ & $47 \pm 4$ & $4400 \pm 700$ & $830 \pm 130$ & $83 \pm 33$ & $590 \pm 190$ & $110 \pm 40$ & $8 \pm 3$\\
\hline
\end{tabular}
\\
$^\mathrm{a}$BF: beam-filling assumed. For the other sources, the size has been adopted as follows: L1157mm and IRAS~4A (extended) from \cite{jor02}; SVS13A (extended) from \cite{lefloch98}; SVS13A (compact) from \cite{looney00}; OMC-2~FIR~4 (extended) from \cite{furlan14}; OMC-2~FIR~4 (compact) from \cite{crim09};  Cep~E from \cite{chini01}; IRAS~4A (compact) from \cite{maret02}; IRAS~16293 from \cite{ali14}.\\
$^\mathrm{b}$References: $^1$\cite{crapsi05}, $^2$\cite{mae99}, $^3$\cite{daniel13}, $^4$\cite{parise14}, $^5$\cite{jor02} $^{6}$\cite{lefloch98}, $^7$\cite{looney00}, $^8$\cite{crim09}, $^9$\cite{crim10a}, $^{10}$\cite{maret02}, $^{11}$\cite{crim10b}.\\
$^\mathrm{c}$For the non-detections of NH$_2$CHO, we have computed a 3$\sigma$ upper limit to its column density adopting the same $T_\mathrm{rot}$ derived for HNCO (see text).\\
$^\mathrm{d}$Data for NH$_2$CHO upper limit derived from 3-mm data by N. Marcelino.\\
$^\mathrm{e}N$(HNCO) is probably a lower limit due to contamination from the OFF position.
\label{trd}
\end{minipage}
\end{table*}

As for the five objects where formamide was not detected, we determined a 3$\sigma$ upper limit to its column density under the assumption of LTE and adopting the corresponding value of $T_\mathrm{rot}$ derived for HNCO. To this end, we used the spectral data around the NH$_2$CHO 4$_{0,4}$ -- 3$_{0,3}$ transition at 84.542~GHz, expected to be the most intense at the cold temperatures implied by the HNCO results. The upper limits thus derived are shown in Table~\ref{trd}.

Once the column densities of HNCO and NH$_2$CHO were obtained, we derived their respective abundances with respect to molecular hydrogen (H$_2$) using the H$_2$ column densities, $N_\mathrm{H2}$, listed in Table~\ref{trd}, which correspond to the indicated source sizes. The uncertainty on $N_\mathrm{H2}$ is included for those sources where this was provided in the corresponding bibliographic reference. The resulting abundances span more than two orders of magnitude and are shown in Table~\ref{trd}, together with their ratio, $R = X(\mathrm{HNCO})/X(\mathrm{NH_2CHO})$.

\subsubsection{Radiative transfer analysis taking into account the source structure}\label{grapes}

The source structures of I16293, IRAS~4A, and OMC-2, are reported in the literature (\citealp{maret02}, \citealp{crim10b}). Therefore, for these objects, a more sophisticated radiative transfer analysis is possible that takes into account the temperature and gas density as a function of distance to the central protostar. Cep~E also has a known structure (\citealp{crim10a}), but having only 5 line detections, both in HNCO and NH$_2$CHO, we do not consider it here.
We have analysed  the I16293, IRAS~4A, and OMC-2 lines by means of the code GRAPES  (GRenoble Analysis of Protostellar Envelope Spectra), whose details are described in \cite{cc03} and \cite{ali14}.

Briefly, GRAPES computes the Spectral Line Energy Distribution (SLED) of a free-infalling spherical envelope with given gas and dust density and temperature profiles, and for a given mass of the central object. The dust to gas ratio is assumed to be the standard one, 0.01 in mass, and the grains have an average diameter of 0.1 $\mu$m. The species abundance is assumed to follow a step-function, with a jump at the dust temperature $T_\mathrm{jump}$, which simulates the thermal desorption of species from icy mantles (e.g. \citealp{cc00}). The abundance $X_\mathrm{i}$ in the warm ($T \geq T_\mathrm{jump}$) envelope is constant. In the outer envelope, we assumed that the abundance follows a power law as a function of the radius, $X_\mathrm{o} r^a$, with an index equal to 0, --1 and --2, as in \cite{ali14}. $X_\mathrm{i}$ and $X_\mathrm{o}$ are considered parameters of the model. Since, to our knowledge, the binding energy of NH$_2$CHO is not available in the literature, we treat $T_\mathrm{jump}$ as a parameter too. However, if the molecules are trapped in water ice, the binding energy of H$_2$O will largely determine the dust temperature at which NH$_2$CHO is injected into the gas phase.

The radiative transfer is solved with the escape probability formalism and the escape probability is computed integrating each line opacity over the 4$\pi$ solid angle. 
We ran models assuming LTE populations for formamide and, for comparison with Sect.~\ref{rot}, HNCO, and models taking into account non-LTE effects for HNCO. In the latter case, we used the collisional coefficients by \cite{green86}, retrieved from the LAMDA database (\citealp{lamda}).

For each molecule and source, we ran a large grid of models varying the four parameters above: $X_\mathrm{i}$, $X_\mathrm{o}$, $T_\mathrm{jump}$, and $a$. In total, we ran about 20,000 models per source. The computed SLED of each model was then compared with the observed SLED to find the solution with the best fit. The results of this analysis are reported in Table~\ref{tgrapes}, where we give the best fit values and the range of $X_\mathrm{i}$, $X_\mathrm{o}$ , $T_\mathrm{jump}$ with $\chi^2\leq1$. We note that the there is no appreciable difference in the best $\chi^2$ when using a different value of $a$, so we took the simplest solution: $a=0$. In this respect, the situation is similar to what \cite{ali14} found in their study of IRAS16293.

\begin{table}
 \centering
  \caption{Results of grapes analysis for NH$_2$CHO and HNCO considering the source structure of IRAS~4A, I16293 and OMC2$^\ast$}
  \begin{tabular}{@{}lccc@{}}
  \hline
  & IRAS~4A & I16293 & OMC-2\\
\hline
\multicolumn{4}{c}{HNCO LTE}\\
\hline
$X_\mathrm{o}$ ($10^{-11}$) & $3 \pm 1$ & $0.1 \pm 0.1$ & $5.5 \pm 1.5$\\
$X_\mathrm{i}$ ($10^{-11}$) & $20 \pm 10$ & $90 \pm 10$ & $< 170$\\
$T_\mathrm{jump}$ (K) & 100 & 40 & 80\\
$T_\mathrm{jump}$~range (K) & 60 -- 120 & 30 -- 50 & $\geq 30$\\
$\chi^2$ & 1.2 & 2.0 & 1.0\\
\hline
\multicolumn{4}{c}{HNCO non-LTE}\\
\hline
$X_\mathrm{o}$ ($10^{-11}$) & $3 \pm 1$ & $0.5 \pm 0.4$ & $4 \pm 1$\\
$X_\mathrm{i}$ ($10^{-11}$) & $30 \pm 20$ & $600 \pm 300$ & $< 20$\\
$T_\mathrm{jump}$ (K) & 100 & 90 & 80\\
$T_\mathrm{jump}$~range (K) & $\geq 50$ & $\geq 60$ & $\geq 30$\\
$\chi^2$ & 1.0 & 1.5 & 0.7\\
\hline
\multicolumn{4}{c}{NH$_2$CHO}\\
\hline
$X_\mathrm{o}$ ($10^{-11}$) & $2 \pm 1$ & $0.3 \pm 0.2$ & $0.3 \pm 0.3$\\
$X_\mathrm{i}$ ($10^{-11}$) & $50 \pm 10$ & $60 \pm 20$ & $200 \pm 50$\\
$T_\mathrm{jump}$ (K) & 100 & 90 & 80\\
$T_\mathrm{jump}$~range (K) & $\geq 100$ & $\geq 50$ & 60 -- 100\\
$\chi^2$ & 2.0 & 0.7 & 1.3\\
\hline
\multicolumn{4}{c}{$R = X(\mathrm{HNCO})/X(\mathrm{NH_2CHO})$}\\
\hline
$R_\mathrm{o}$ (LTE) & $1.5 \pm 0.9$ & $< 1.7$ & $18 \pm 18$\\
$R_\mathrm{i}$ (LTE) & $0.4 \pm 0.2$ & $1.5 \pm 0.5$ & $< 0.85$\\
$R_\mathrm{o}$ (non-LTE) & $1.5 \pm 0.9$ & $1.7 \pm 1.7$ & $13 \pm 13$\\
$R_\mathrm{i}$ (non-LTE) & $0.6 \pm 0.4$ & $10 \pm 6$ & $< 0.1$\\
\hline
\end{tabular}
\\
$^\ast$Abundances with respect to H$_2$ are times 10$^{-11}$. $X_\mathrm{o}$ and $X_\mathrm{i}$ are the outer and inner abundances, respectively.\\
\label{tgrapes}
\end{table} 

A comparison between the results obtained for HNCO with the LTE and non-LTE level populations shows that the LTE approximation is quite good in the case of IRAS~4A and OMC-2, but not for I16293. The reason for that is probably a lower density envelope of I16293 compared to the other two sources. Therefore, the LTE results are likely reliable also for the formamide in IRAS 4A and OMC-2, while in I16293 these have to be taken with some more caution.

A second result of the GRAPES analysis is that both HNCO and formamide have a jump in their abundances at roughly the same dust temperature, 80--100 K. This is an important result reflecting the two molecules have similar behaviours with changes in temperature. It suggests they trace the same regions within the analysed protostars.

In order to evaluate whether the rotational diagram (hereafter RD) and GRAPES analyses are in agreement, we compare their respective abundance values, which roughly agree within an order of magnitude, in the Appendix~\ref{comparison}. We note here that, while the GRAPES analysis is likely more accurate, we are not able to apply it to the other sources of this study due to the lack of known source structure and/or lack of a sufficient amount of molecular lines. The absence of interferometric imaging of the HNCO and NH$_2$CHO emission also hinders the study of the inner structure of the protostellar emission. Therefore, we base the discussion below largely on the RD results, with a note of caution that those values may not strictly represent the physical properties of the sources.

\section{Discussion}\label{disc}

\subsection{Formation routes of NH$_2$CHO}\label{routes}

The formation mechanism(s) of interstellar formamide, as that of other COMs, is still far from being established. Several routes have been proposed so far which include both gas-phase and grain-surface processes. Concerning the former, \cite{qh07} suggested NH$_2$CHO forms via the radiative association reaction

\begin{equation}
\mathrm{NH_4^+ + H_2CO} \rightarrow \mathrm{NH_4CH_2O^+ + h\nu}
\label{eqh}
\end{equation}
followed by dissociative recombination. \cite{halfen11} proposed the following ion-molecule reaction and subsequent electron recombination:
\begin{equation}
\mathrm{NH_4^+ + H_2CO} \rightarrow \mathrm{NH_3CHO^+ + H_2}
\label{ehalfen1}
\end{equation}
\begin{equation}
\mathrm{NH_3CHO^+ + e^-} \rightarrow \mathrm{NH_2CHO + H}
\label{ehalfen2}
\end{equation}
These reactions all have unknown rates. Thus, further experimental work will be needed in order to evaluate their effectiveness in producing formamide.

Neutral neutral reactions have also been discussed as possible gas-phase routes leading to NH$_2$CHO. In particular, \cite{garrod08} proposed the radical-neutral reaction

\begin{equation}
\mathrm{H_2CO + NH_2} \rightarrow \mathrm{NH_2CHO + H}
\label{egarrod}
\end{equation}
However, as recently mentioned by \cite{redondo14}, it presents a net activation barrier of $>1000$~K that makes it inviable in interstellar conditions. Other neutral-neutral reactions evaluated by these authors also revealed to have large activation barriers, thus ruling them out as dominant or efficient mechanisms to produce NH$_2$CHO.

Formamide may also be formed on the icy mantles of dust grains. \cite{jones11} conducted some experimental work in which they irradiate a mixture of ammonia (NH$_3$) and carbon monoxide (CO) ices with high-energy (keV) electrons, resulting in NH$_2$CHO as one of the final products. The authors discuss several possible reactions and conclude that the most plausible route towards formamide begins with the cleavage of the nitrogen-hydrogen bond of ammonia, forming the NH$_2$ radical and atomic H. The latter, containing excess kinetic energy, can then add to CO, overcoming the entrance barrier, to produce the formyl radical (HCO). Finally, HCO can combine with NH$_2$ to yield NH$_2$CHO.

A different grain-mantle mechanism was proposed by \cite{garrod08}, who considered hydrogenation (i.e. addition of H atoms) of OCN in their chemical models. However, this route resulted in an overabundance of NH$_2$CHO and an underabundance HNCO, since the latter was efficiently hydrogenated to yield formamide, the final product. \cite{raunier04} performed experimental Vacuum Ultra Violet (VUV) irradiation of solid HNCO at 10~K, which led to NH$_2$CHO among the final products. They proposed that photodissociation of HNCO yields free H atoms that subsequently hydrogenate other HNCO molecules in the solid to finally give NH$_2$CHO. The limitation of this experiment is that it was carried out with pure solid HNCO. \cite{jones11} mentioned that, in the presence of NH$_3$, quite abundant in grain mantles, HNCO will preferentially react with it, resulting in NH$_4^+ +$~OCN$^-$. Despite these caveats, hydrogenation of HNCO on grain mantles was recently found to be a most likely solution in the case of the outflow shock regions L1157-B1 and B2 (\citealp{edgar14}). More experiments and calculations are needed in order to assess the efficiency of this formation route.

\subsection{Correlation between HNCO and NH$_2$CHO}

From the previous section, it is clear that, until more gas-phase and surface reaction rates involving the mentioned species are measured, it will be difficult to establish the exact synthesis mechanisms of formamide in space.

\begin{figure}
\centering
\begin{tabular}{c}
  \includegraphics[scale=0.48]{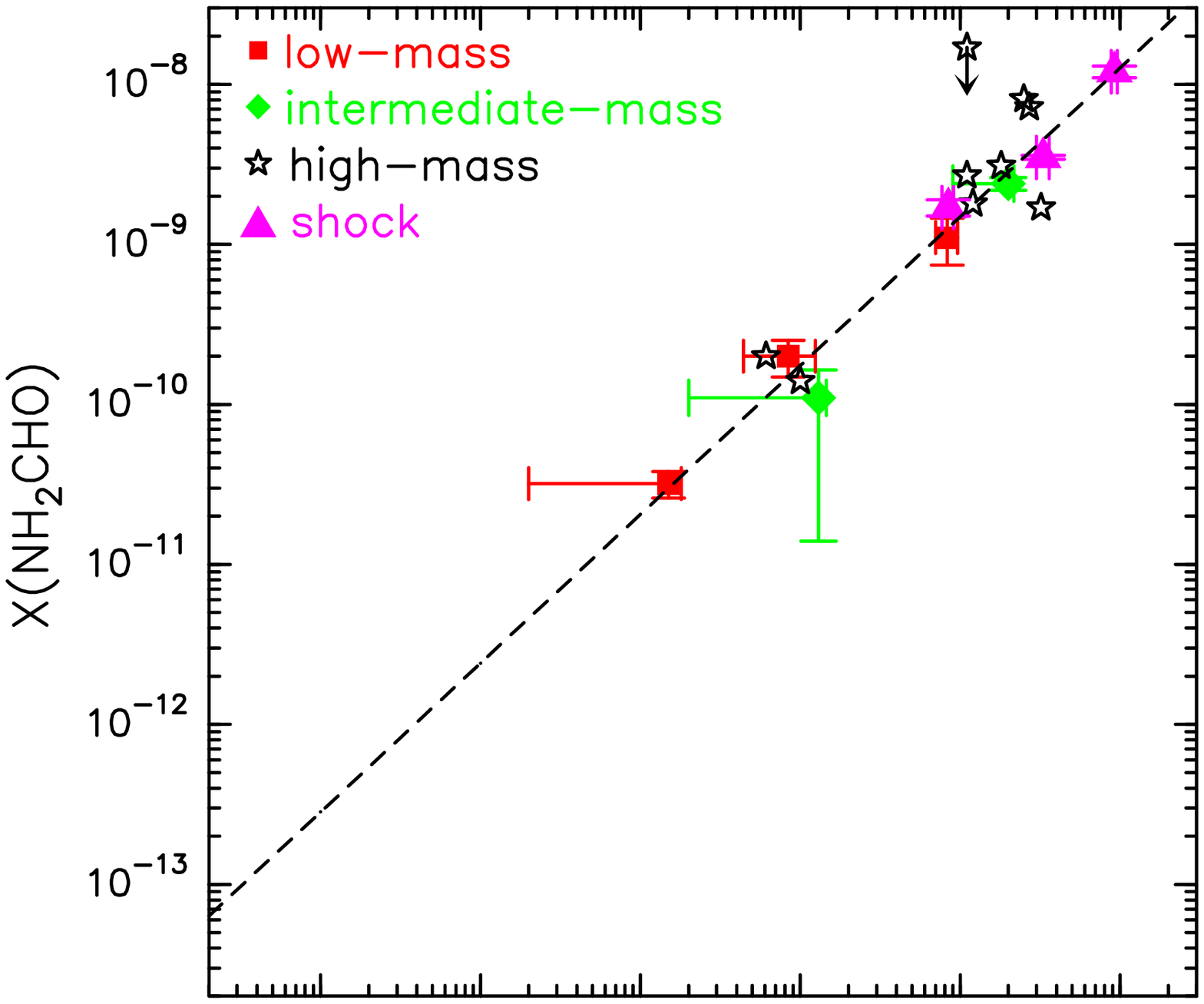} \\
  \includegraphics[scale=0.48]{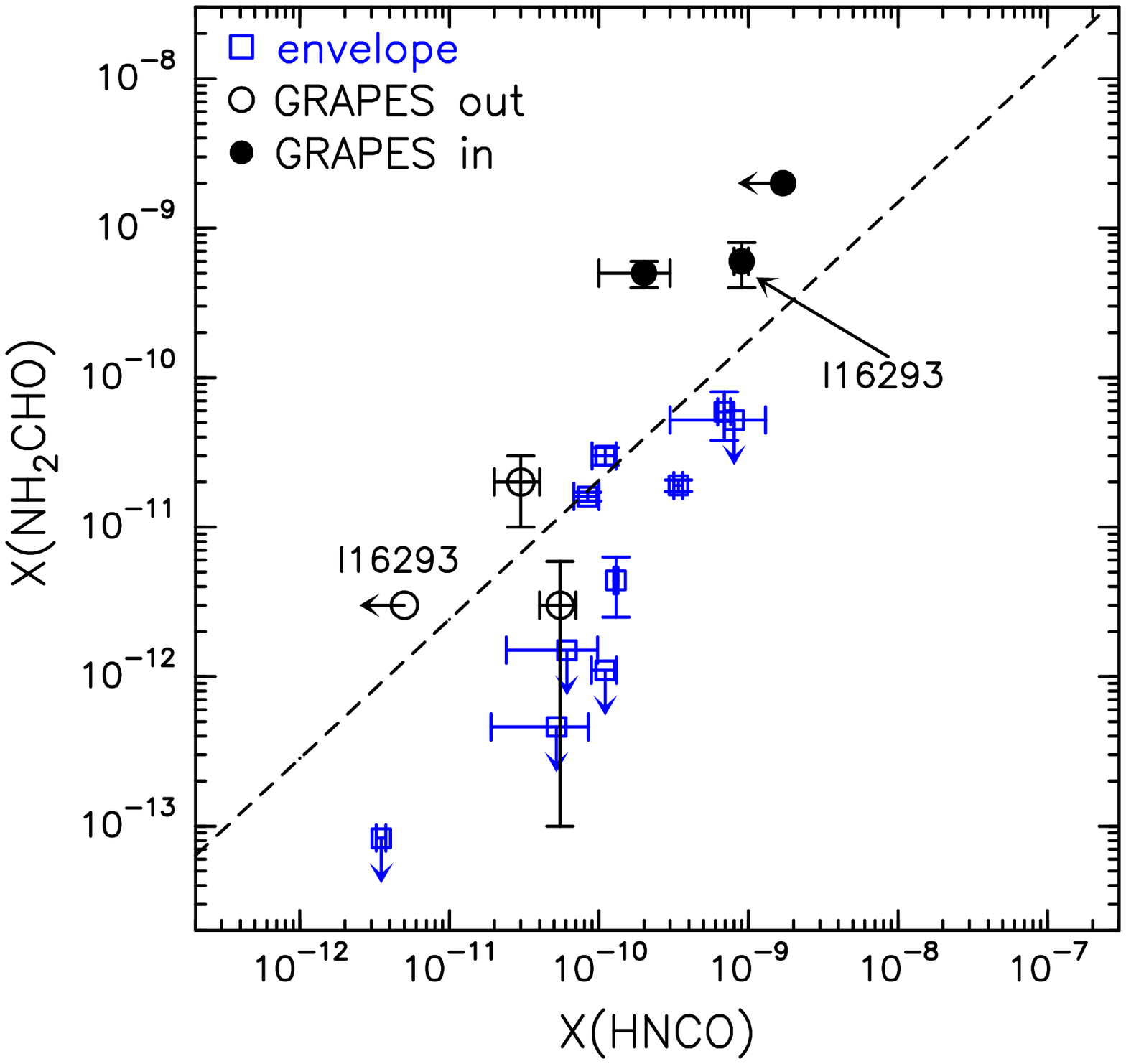} \\ 
\end{tabular}
  \caption{Plot of NH$_2$CHO versus HNCO abundances with respect to H$_2$. \textit{Top}: Data points included in the power-law fit (dashed line; see text). Red squares and green diamonds denote the compact or inner RD solutions of low- and intermediate-mass sources in this study, respectively. Magenta triangles and black stars correspond, respectively, to outflow shock regions (from \citealp{edgar14}) and high-mass sources (from \citealp{bis07} and \citealp{num00}). \textit{Bottom}: Data points not included in the power-law fit (see text). Blue open squares represent the extended or outer RD solutions, while black open and filled circles denote the GRAPES LTE values for the outer and inner components, respectively.}
  \label{fabu}
\end{figure}

In this section, we assess, from an observational point of view, whether hydrogenation of HNCO leading to NH$_2$CHO on the icy mantles of dust grains could be a dominant formation route. To this aim, we plot in Fig.~\ref{fabu}, the NH$_2$CHO versus HNCO abundances of all our sources (Table~\ref{trd}), as well as the shock regions analysed by \cite{edgar14}, and the high-mass SFRs reported in \cite{bis07} and \cite{num00}, for comparison. The latter were obtained by the cited authors via the RD method assuming the emission comes from the inner hot core regions. Thus, for homogeneity, we split the plot into two panels, the upper one showing only the compact/inner solutions of the RD analysis, classified by masses. The best power law fit to these points is marked with a dashed line, and is given by the equation $X(\mathrm{NH_2CHO}) = 0.04  X(\mathrm{HNCO})^{0.93}$, with a Pearson coefficient of 0.96, indicating a tight correlation. The fact that this correlation is almost linear and holds for more than three orders of magnitude in abundance suggests that HNCO and NH$_2$CHO are chemically related. This result confirms, on a more statistical basis, what was recently found by \cite{edgar14}. 

However, this correlation does not hold for the objects without formamide detections, which are plotted in the lower panel of Fig.~\ref{fabu} together with the extended envelope solutions of the RD analysis. Here, it is clearly seen that all the upper limits lie well below the best fit line, indicating a significantly larger amount of gas-phase HNCO relative to NH$_2$CHO in comparison to the other sources. These objects are the coldest in our sample, representing either pre-stellar cores or protostars with no detectable hot corino within them. The rotational temperatures inferred from the HNCO RD analysis are also among the lowest in our sample. In this same plot, the points representing formamide detections (extended envelope component) also show a tendency towards lower relative values of $X$(NH$_2$CHO), although not as pronounced.

Thus, it appears that regions with colder temperatures are more deficient in NH$_2$CHO than protostars with hot inner regions, indicating that higher temperatures are needed for NH$_2$CHO to become relatively abundant in the gas phase. This might be explained by (i) NH$_2$CHO forming in the gas phase at temperatures above $\sim 100$~K, and/or (ii) it forming predominantly on the icy mantles of dust grains at low temperatures, and subsequently sublimating into the gas-phase when the temperature in the inner regions rises sufficiently. As for the former possibility, \cite{edgar14} quantitatively argued that reaction~\ref{egarrod} does not suffice to explain the amount of gas-phase formamide in the shock regions of L1157 protostellar outflow. In addition, the high activation barrier the reaction needs to overcome makes this an unviable route. Other purely gas-phase formation routes still need more investigation in terms of reaction rates and activation barriers, as discussed in Sect.~\ref{routes}. \cite{edgar14} favoured a grain formation mechanism followed by mantle-grain evaporation/sputtering on the basis of the comparable abundance enhancements of HNCO, NH$_2$CHO, and CH$_3$OH in the gas-phase between the two protostellar shocks studied by the authors. Therefore, grain formation of NH$_2$CHO appears to be the most likely possibility.

On the other hand, while grain formation of HNCO is likely to occur in the cold phases of star formation (\citealp{hase93}), gas-phase reactions leading to HNCO at such cold temperatures can also take place efficiently (see e.g. \citealp{marcelino09} and references therein), overcoming strong depletion. This would explain its relatively high gas-phase abundance already in the very early --and cold-- phases of star formation, and also the high values of HNCO to NH$_2$CHO abundance ratios we find in the coldest sources of our sample.

\begin{figure}
\centering
\begin{tabular}{c}
  \includegraphics[scale=0.48]{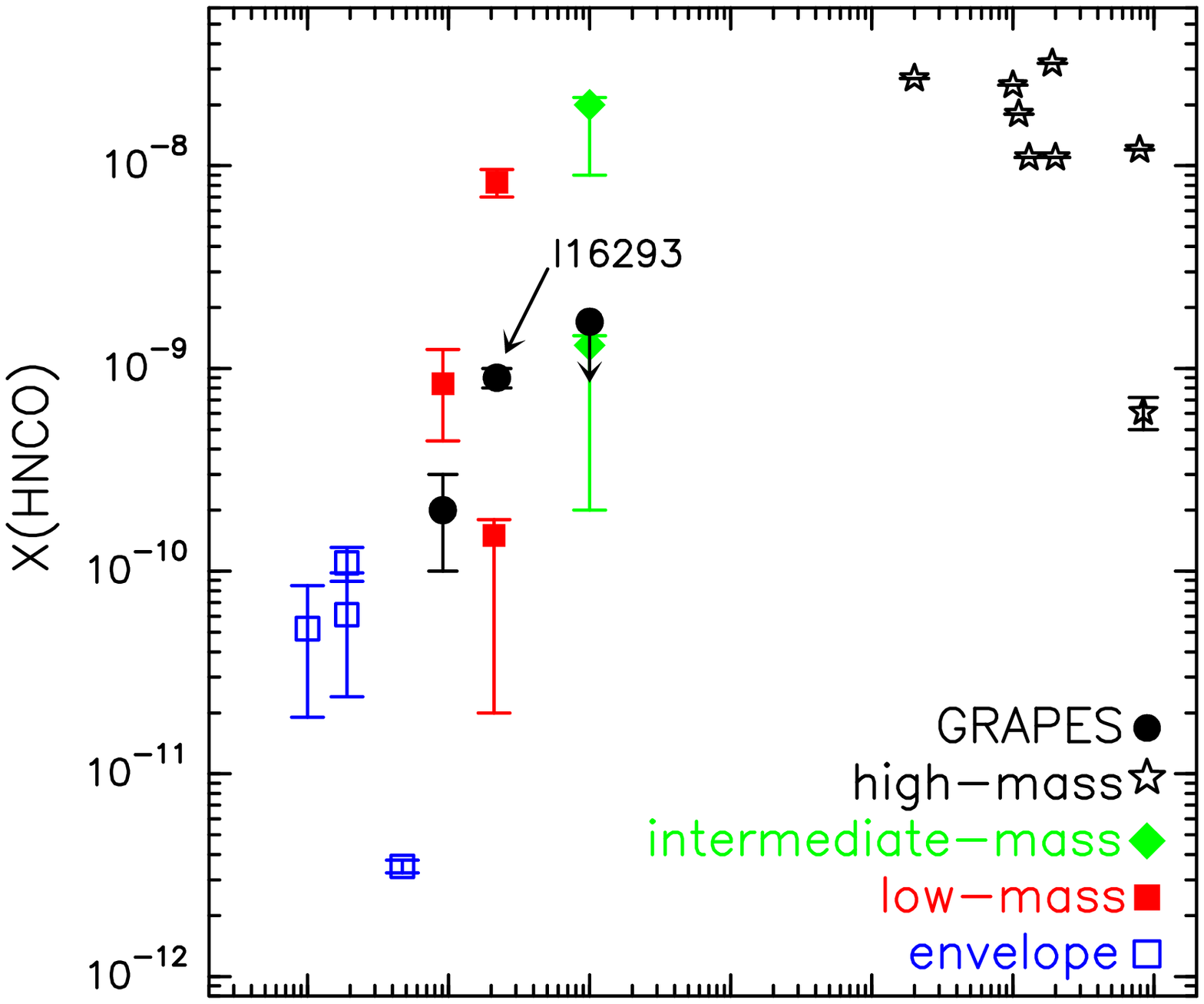}\\
  \includegraphics[scale=0.48]{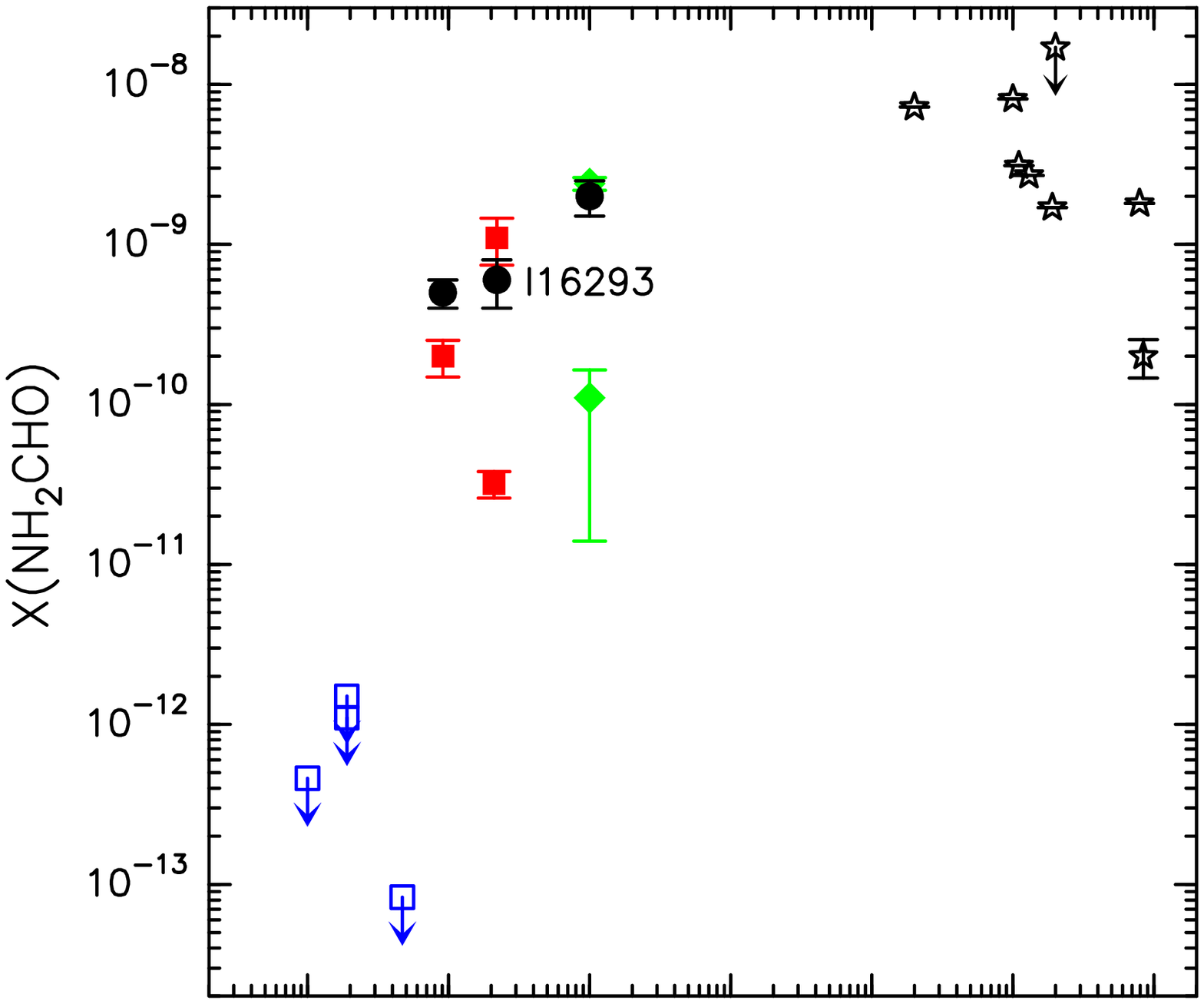}\\
  \includegraphics[scale=0.48]{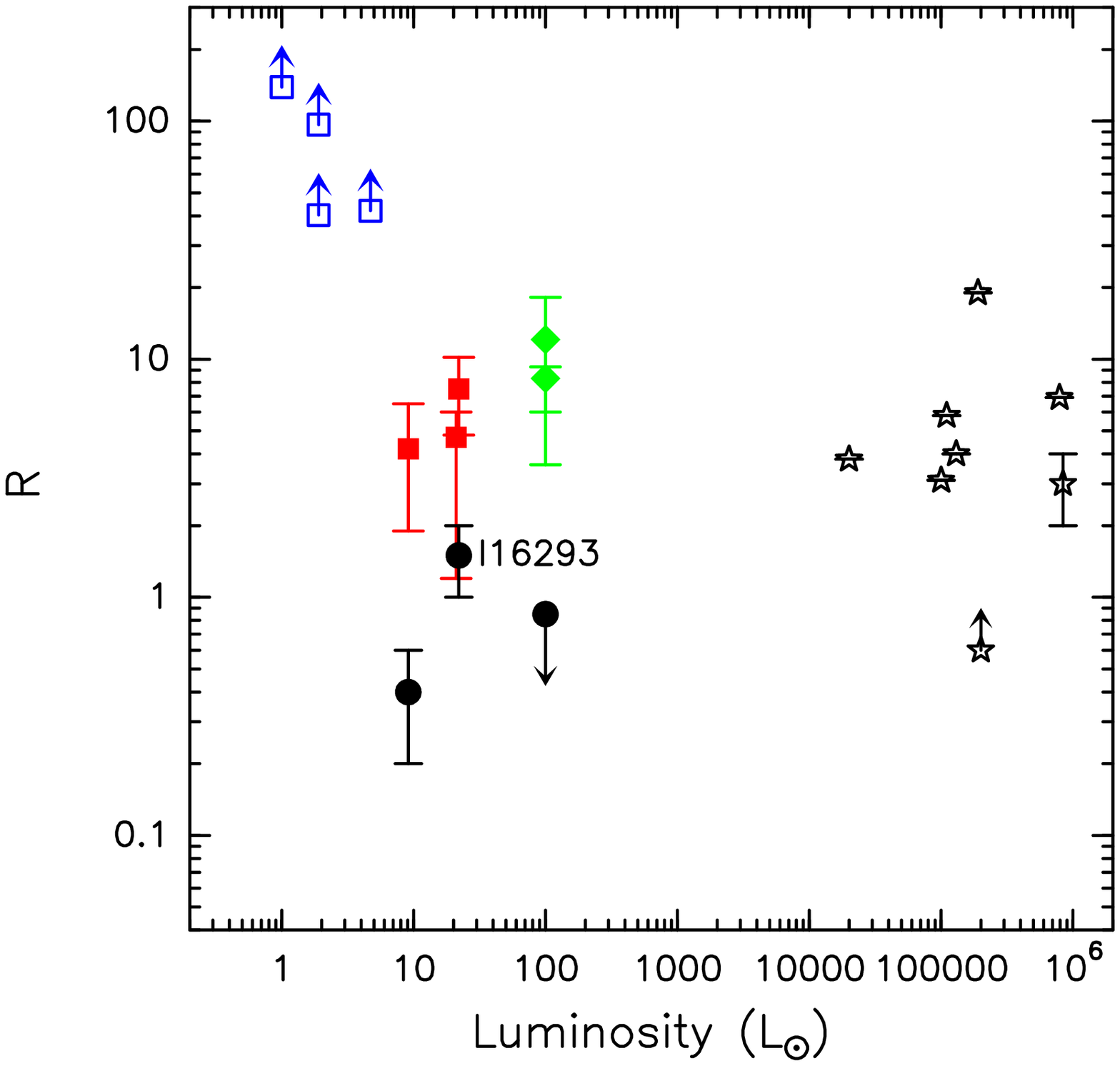}\\
\end{tabular}
  \caption{Abundance of HNCO (\textit{top}), NH$_2$CHO (\textit{middle}) and their ratio (\textit{bottom}) against bolometric luminosity. Symbols are as in Fig.~\ref{fabu}.}
  \label{flum}
\end{figure}

In Fig.~\ref{flum}, we plot the HNCO abundance, the NH$_2$CHO abundance, and their ratio, $R$, as a function of bolometric luminosity for those sources with a reported luminosity estimate (see Table~\ref{tsample}). For the objects in our study with formamide detection, we only plot the points corresponding to the inner or compact component (red circles), since these regions are expected to be the dominant contributors to the overall luminosity. The HNCO and NH$_2$CHO abundance panels both show the high-mass sources lying on top of the plot, while the points representing our sample sources are more scattered, with the coldest objects (in blue) showing the lowest abundances. This trend is much more pronounced in the case of NH$_2$CHO, for which hot corino regions (red points) display higher NH$_2$CHO abundances than the colder objects by more than an order of magnitude. More interesting is the plot of $R$, which illustrates how this quantity remains roughly constant along 6 orders of magnitude in luminosity for the NH$_2$CHO-emitting sources, with values ranging from 3 to 10 approximately. This reflects the almost-linearity of the correlation between the abundance of the two species. On the other hand, this value rises considerably for the lower luminosity sources, re-enforcing our interpretation that formamide mostly forms on grains at cold temperatures, while HNCO may form both on grains and in the gas.

The strikingly tight and almost linear correlation between the abundance of the two molecules once NH$_2$CHO becomes detectable suggests one of the two following possibilities: (i) HNCO and NH$_2$CHO are both formed from the same parent species on dust grain mantles, or (ii) one forms from the other. Among the grain formation routes that have been proposed so far, hydrogenation of HNCO leading to NH$_2$CHO is the only mechanism that would explain our observational results. While this route is found to have some caveats (see Sect.~\ref{routes}), it is also true that more experimental work is needed to better assess its efficiency.

If the abundance of gaseous NH$_2$CHO truly depends on temperature, we should find a difference in $R$ between the hot corino and the cold envelope regions of IRAS~4A and I16293. Looking at Table~\ref{trd}, this is indeed the case. As for OMC-2, Cep~E, and SVS13A, only one component was necessary to describe their rotational temperatures and column densities. Therefore, we cannot compare the extended and compact values as in the case of a 2-component solution. We can nevertheless guess that, excluding the case of Cep~E, for which only  low-energy formamide lines were detected, the compact solution is likely the best, given the low values of $R$ and the relatively high rotational temperatures derived. This would imply most of the emission arises in the inner hot corino regions. For Cep~E, more molecular observations at higher frequencies are needed to confirm this.

Figures~\ref{fabu} and \ref{flum} also include the results from the LTE GRAPES analysis. I16293 (labelled in the plots) is included for completeness despite the fact that the GRAPES analysis suggests non-LTE effects should be taken into account for this object. While these points introduce more scatter in the plots, it can be clearly seen that the inner components of the sources analysed with GRAPES have a lower HNCO abundance relative to NH$_2$CHO, compared to what is found via the RD analysis. This yields lower $R$ values, indicating a considerable amount of formamide with respect to HNCO in these regions and suggesting, as mentioned in Sect.~\ref{comparison}, that the 2-component approximation in the RD analysis is over simplistic: while we assumed that only the higher-energy formamide lines arose from the compact inner region, it is likely that a significant amount of emission from the low-energy lines also originates here and not exclusively in the outer envelope.

The trend showing higher $R$ in the outer envelope than in the inner regions holds for both IRAS~4A and OMC-2, which further supports the fact that NH$_2$CHO requires higher temperatures than HNCO to be detectable in the gas phase. This kind of analysis, taking into account the source structure, is needed in a larger sample of objects in order to draw conclusions about both the chemistry and the validity of our RD analysis on a more statistical basis. Interferometric mapping would also greatly help disentangling source multiplicity and verifying whether the emission of HNCO and NH$_2$CHO trace the same regions, as has been assumed in this work.

\section{Conclusions}\label{conclusions}

As part of the IRAM Large programme ASAI, we searched for millimetre spectral lines from formamide (NH$_2$CHO), a presumably crucial precursor of pre-biotic material, and isocyanic acid (HNCO), in ten low- and intermediate-mass star forming regions with different properties. The dataset, obtained with the IRAM~30-m telescope, consists mainly of unbiased broadband spectral surveys at 1, 2, and 3-mm. Our aim was to investigate the chemical connection between these two molecular species and gain some observational insights into the formation mechanisms of formamide in interstellar conditions. The present work represents the first systematic study within ASAI and statistically completes the low-mass end of similar studies performed towards high-mass star-forming regions. Our main findings are summarised as follows.

\begin{enumerate}
\item[1.] The high sensitivity and large frequency range of the spectral surveys allowed us to evaluate the detectability of numerous NH$_2$CHO and HNCO transitions. We detect formamide in five out of the ten objects under study (IRAS~4A, IRAS~16293, SVS13A, Cep~E, and OMC-2), and HNCO in all of them. Since formamide had already been detected in IRAS~16293 --also investigated here for completeness--, this study raises the number of known low- and intermediate-mass formamide-emitting protostars to five, thus significantly improving the statistics.
\item[2.] We derived HNCO and NH$_2$CHO column densities via the rotational diagram method for all the sources. As a result, we found NH$_2$CHO abundances with respect to H$_2$ in the range 10$^{-11}$ -- 10$^{-9}$, and HNCO abundances between 10$^{-12}$ and 10$^{-8}$. For those objects without formamide detection, we provided an upper limit to its column density and abundance.
\item[3.] For three targets (IRAS~4A, IRAS~16293, and OMC-2), the source density and temperature structures are known and published, and we were thus able to take them into account through a more sophisticated analysis using the code GRAPES. This method fits an abundance profile that consists of a step function, with the separation between the two values roughly corresponding to the hot corino size. A comparison between the two radiative transfer analyses employed reveals overall agreement within an order of magnitude. The GRAPES analysis also indicates that one of the studied objects, IRAS~16293, requires a non-LTE radiative transfer analysis, which at the moment is not possible due to the lack of collisional coefficients for NH$_2$CHO. LTE appears to describe correctly the other two sources analysed with GRAPES, and is assumed to be a good approximation for all the other sources in our sample.
\item[4.] For the sources where formamide was detected, i.e. hot corino sources, we found an almost linear correlation between HNCO and NH$_2$CHO abundances that holds for several orders of magnitude. This suggests that the two molecules may be chemically associated. On the other hand, those sources with no formamide detection do not follow this correlation, but instead show much larger amounts of HNCO relative to NH$_2$CHO. These objects are the coldest in this study, and unlike the rest of our sample, they contain no known hot corinos.
\item[5.] Our findings and the NH$_2$CHO formation routes proposed so far in the literature suggest that, unlike HNCO, NH$_2$CHO does not form efficiently in the gas phase at cold temperatures and may be formed on the mantles of dust grains, where it remains frozen at cold temperatures. As soon as the temperature rises sufficiently to sublimate the icy grain mantles, formamide is incorporated into the gas and becomes detectable. The tight and almost linear correlation with HNCO suggests a possible formation route of NH$_2$CHO via hydrogenation of HNCO, although other possibilities should not be ruled out. In particular, two potentially viable gas-phase pathways leading to formamide involve formaldehyde (H$_2$CO). It is therefore worth exploring the connection between H$_2$CO and NH$_2$CHO, which will be the subject of a forthcoming paper.
\item[6.] In order to evaluate the validity of our conclusions, several aspects need to be explored more thoroughly. From an observational point of view, interferometric imaging is necessary to assess the relative spatial distribution of HNCO and NH$_2$CHO, and retrieve more accurate abundance ratios, in particular in the hot corino sources. In addition, more detailed and sophisticated radiative transfer analysis requires, on the one hand, knowledge of the source density and temperature profiles, and on the other hand, collisional coefficient calculations for NH$_2$CHO, currently unavailable. Finally, more chemical experiments are needed to estimate the efficiency of the hydrogenation processes leading from isocyanic acid to formamide on interstellar dust grains, as well as the viability of purely gas-phase reactions.
\end{enumerate}

\section*{Acknowledgments}

We would like to thank the staff members at IRAM, who greatly helped before, during, and after the ASAI observations. We are also grateful to our anonymous referee, whose comments helped us to improve the manuscript, and to Y. Watanabe, N. Sakai, and S. Yamamoto for very useful discussions. A.L.S. acknowledges financial support from Grant-in-Aids from the Ministry of Education, Culture, Sports, Science, and Technologies of Japan (25108005). This work is partly supported by the French Space Agency CNES (Centre National d'\'Etudes Spatiales) and the PRIN INAF 2012 -- JEDI and by the Italian Ministero dell'Istruzione, Universit\`a e Ricerca through the grant Progetti Premiali 2012 -- iALMA. MK acknowledges support from a Royal Netherlands Academy of Arts and Sciences (KNAW) professor prize. M.T. and R.B. gratefully acknowledge partial support from MINECO Grant FIS2012-32096.

\appendix

\section{Comparison between GRAPES and rotational diagram analyses}\label{comparison}

This section aims to compare the agreement between the rotational diagram and GRAPES methods. As described in Sect~\ref{grapes}, the line emission in I16293 does not appear to be well described by LTE, and a more realistic radiative transfer treatment will need to wait until collisional coefficients are available for NH$_2$CHO. Therefore, we do not consider it here, while it is worth noticing that a rotational diagram analysis is likely too simplistic to analyse the HNCO and NH$_2$CHO lines in this source.

In OMC-2, the GRAPES analysis tells us that the temperature that separates the inner and outer components is 80~K, both in the LTE and non-LTE approximations. Thus, for consistency in the comparison, we re-computed the inner and outer abundances resulting from the RD analysis using the same inner sizes as in GRAPES, instead of those corresponding to a temperature of the 100~K (see Sect.~\ref{rot}). We note that, while the RD analysis allowed for a separation of two components (inner and outer) for IRAS~4A, a single component was sufficient for OMC-2. It should be kept in mind, therefore, that for the latter the comparison is not equivalent, since we are not comparing a two-component solution with another two-component solution as in the case of the other two protostars.

\begin{table}
 \centering
  \caption{Comparison between GRAPES and RD analyses$^\ast$}
  \begin{tabular}{@{}lcc@{}}
  \hline
  & IRAS~4A & OMC-2\\
\hline
Inner size ($''$) & 1.5 & 3.1\\
\hline
\multicolumn{3}{c}{RD-to-GRAPES ratio (LTE)}\\
\hline
$X_\mathrm{o}$(HNCO) & $11 \pm 4$ & $1.5 \pm 0.5$\\
$X_\mathrm{i}$(HNCO) & $4 \pm 3$ & $> 2.5$\\
$X_\mathrm{o}$(NH$_2$CHO) & $1 \pm 1$ & $5 \pm 5$\\
$X_\mathrm{i}$(NH$_2$CHO) & $0.4 \pm 0.4$ & $0.3 \pm 0.3$\\
$R_\mathrm{o}$ & $12 \pm 7$ & $0.3 \pm 0.3$\\
$R_\mathrm{i}$ & $11 \pm 8$ & $> 10$\\
\hline
\multicolumn{3}{c}{RD-to-GRAPES ratio (non-LTE)}\\
\hline
$X_\mathrm{o}$(HNCO) & $11 \pm 4$ & $2.1 \pm 0.7$\\
$X_\mathrm{i}$(HNCO) & $3 \pm 2$ & $> 22$\\
$X_\mathrm{o}$(NH$_2$CHO) & $1 \pm 1$  & $5 \pm 5$\\
$X_\mathrm{i}$(NH$_2$CHO) & $0.4 \pm 0.4$ &$0.3 \pm 0.3$\\
$R_\mathrm{o}$ & $12 \pm 7$ & $0.4 \pm 0.4$\\
$R_\mathrm{i}$ & $7 \pm 6$ & $> 85$\\
\hline
\end{tabular}
\\
$^\ast$Abundances with respect to H$_2$ are times 10$^{-11}$.\\
\label{tcomp}
\end{table}

\begin{figure}
\centering
  \includegraphics[scale=0.63]{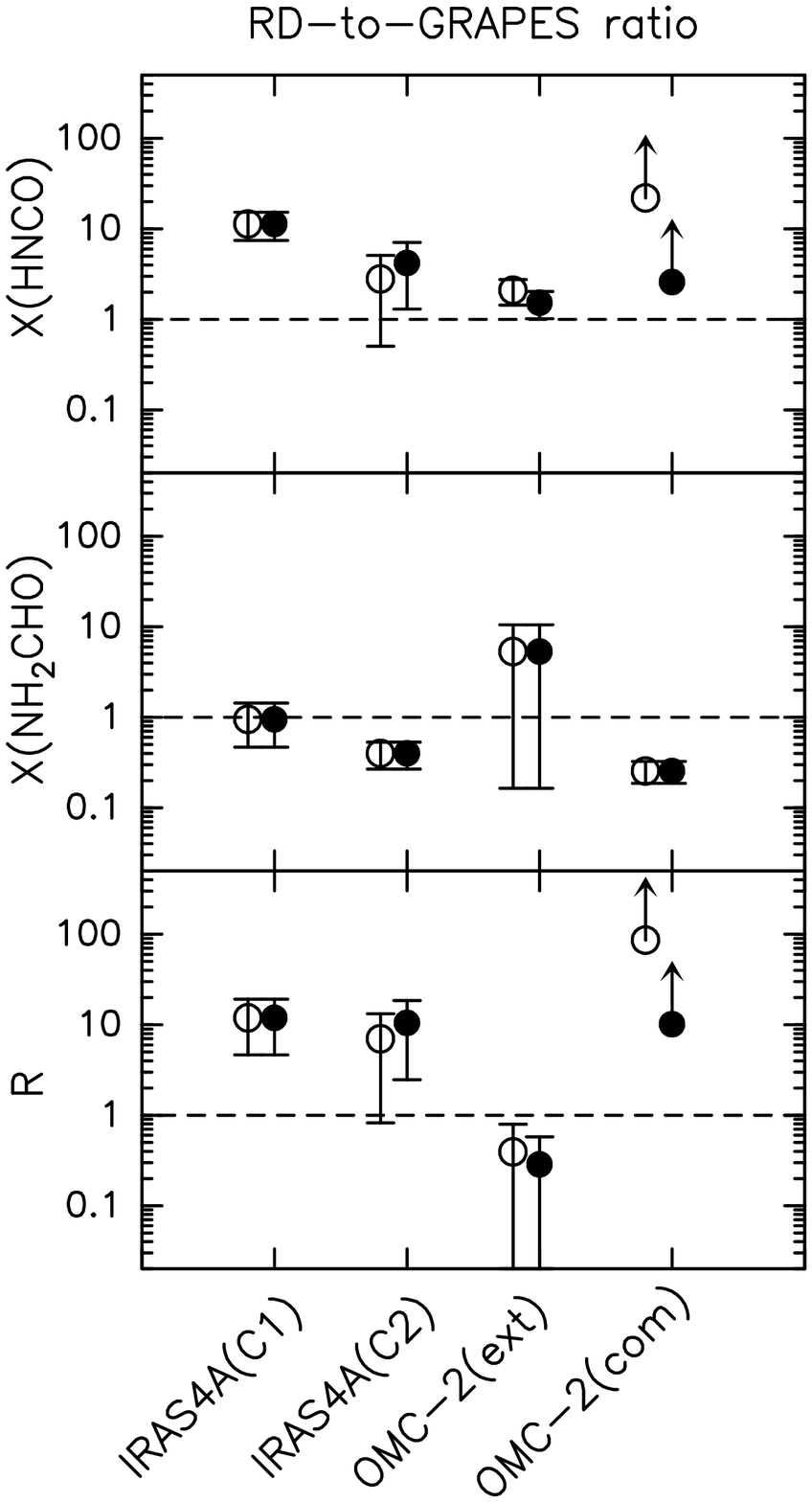} 
  \caption{Ratio of RD-to-GRAPES abundances. \textit{Top}: HNCO abundance. \textit{Middle}: NH$_2$CHO abundance. \textit{Bottom}: HNCO to NH$_2$CHO abundance ratio. Filled and open circles represent, respectively, the LTE and non-LTE HNCO solution in the GRAPES analysis. The horizontal dashed lines mark equality between RD and GRAPES values.}
  \label{fcomp}
\end{figure}

The results of the comparison are listed in Table~\ref{tcomp} and illustrated in Fig.~\ref{fcomp}, where we present the comparison using both the LTE and non-LTE results from GRAPES. It is evident that LTE and non-LTE yield practically the same results for these two sources. It can also be seen that the errors are quite high in some cases, up to 100\%, which are caused by the large uncertainties resulting from the GRAPES analysis. Taking these into account, we find the following behaviours:

\begin{itemize}
\item HNCO abundance: Generally, both methods agree within an order of magnitude, but there is a tendency towards higher values in the RD analysis, by a factor of a few.
\item NH$_2$CHO abundance: Again, we find agreement within a factor of a few. The compact solution is systematically lower in the RD treatment. This suggests that a non-negligible amount of emission from low-energy molecular lines actually comes from the inner region, and not exclusively from the extended envelope, as assumed in the linear fitting of the RD. Such a finding reflects the necessity of analysis like that performed with GRAPES if we want to properly disentangle the inner and outer components in hot corino or hot core sources.
\item HNCO to NH$_2$CHO abundance ratio, $R$: In this case, the two analysis methods agree within a factor of a few.
\end{itemize}

\newpage

\section[]{Tables}

\begin{table*}
\centering
 \begin{minipage}{175mm}
  \caption{NH$_2$CHO transitions searched for in this study and 3$\sigma$ detections$^\mathrm{a}$}
\begin{tabular}{ccccccccccc}
\hline
Transition & $\nu$ & $E_u$ & $A_{ul}$ & $\theta_\mathrm{b}$ & OMC-2 & CepE & SVS13A & IRAS4A & I16293 & Blends\\
 & (MHz) & (K) & ($10^{-5}$~s$^{-1}$) & ($''$) &  &  &  &  &  & \\
\hline
4$_{1,4}$ -- 3$_{1,3}$ & 81693.446 & 12.8 & 3.46 & 30 & N & Y$^\ast$ & N & N & N & \\
4$_{0,4}$ -- 3$_{0,3}$ & 84542.330 & 10.2 & 4.09 & 29 & Y & Y & N & N & Y & \\
4$_{2,3}$ -- 3$_{2,2}$ & 84807.795 & 22.1 & 3.09 & 29 & N & N & N & N & N & \\
4$_{3,2}$ -- 3$_{3,1}$ & 84888.994 & 37.0 & 1.81 & 29 & N & N & N & N & N & \\
4$_{3,1}$ -- 3$_{3,0}$ & 84890.987 & 37.0 & 1.81 & 29 & N & N & N & N & N & \\
4$_{2,2}$ -- 3$_{2,1}$ & 85093.272 & 22.1 & 3.13 & 29 & N & Y & N & N & N & \\
4$_{1,3}$ -- 3$_{1,2}$ & 87848.873 & 13.5 & 4.30 & 28 & Y & Y$^\ast$ & N & N & Y & \\
5$_{1,5}$ -- 4$_{1,4}$ & 102064.267 & 17.7 & 7.06 & 24 & Y & Y & N & B & Y & H$_2$COH$^+$?\\
5$_{0,5}$ -- 4$_{0,4}$ & 105464.219 & 15.2 & 8.11 & 23 & Y & Y & Y & Y & Y & \\
5$_{2,4}$ -- 4$_{2,3}$ & 105972.599 & 27.2 & 6.92 & 23 & Y & N & N & N & N & \\
5$_{4,2}$ -- 4$_{4,1}$ & 106107.870 & 63.0 & 2.98 & 23 & N & N & B & B & N & NH$_2$CHO\\
5$_{4,1}$ -- 4$_{4,0}$ & 106107.895 & 63.0 & 2.98 & 23 & N & N & B & B & N & NH$_2$CHO\\
5$_{3,3}$ -- 4$_{3,2}$ & 106134.427 & 42.1 & 5.29 & 23 & N & N & N & N & N & \\
5$_{3,2}$ -- 4$_{3,1}$ & 106141.400 & 42.1 & 5.29 & 23 & N & N & N & N & N & \\
5$_{2,3}$ -- 4$_{2,2}$ & 106541.680 & 27.2 & 7.03 & 23 & N & N & N & Y & N & \\
5$_{1,4}$ -- 4$_{1,3}$ & 109753.503 & 18.8 & 8.78 & 22 & N & N & N & N & N & \\
\hline
6$_{1,5}$ -- 5$_{1,4}$ & 131617.902 & 25.1 & 15.6 & 19 & Y & --- & N & N & N & \\
7$_{1,7}$ -- 6$_{1,6}$ & 142701.325 & 30.4 & 20.2 & 17 & Y & --- & Y & Y & N & \\
7$_{0,7}$ -- 6$_{0,6}$ & 146871.475 & 28.3 & 22.5 & 17 & B & --- & B & B & B & CH$_3$OCH$_3$\\
7$_{2,6}$ -- 6$_{2,5}$ & 148223.143 & 40.4 & 21.2 & 17 & B & --- & N & N & B & HCNH$^+$\\
7$_{6,1}$ -- 6$_{6,0}$ & 148555.852 & 135.7 & 6.18 & 17 & N & --- & N & N & N & \\
7$_{6,2}$ -- 6$_{6,1}$ & 148555.852 & 135.7 & 6.18 & 17 & N & --- & N & N & N & \\
7$_{5,3}$ -- 6$_{5,2}$ & 148566.822 & 103.0 & 11.4 & 17 & N & --- & N & N & N & \\
7$_{5,2}$ -- 6$_{5,1}$ & 148566.823 & 103.0 & 11.4 & 17 & N & --- & N & N & N & \\
7$_{4,4}$ -- 6$_{4,3}$ & 148598.970 & 76.2 & 15.7 & 17 & N & --- & N & N & N & \\
7$_{4,3}$ -- 6$_{4,2}$ & 148599.354 & 76.2 & 15.7 & 17 & N & --- & N & N & N & \\
7$_{3,5}$ -- 6$_{3,4}$ & 148667.301 & 55.3 & 19.1 & 17 & Y & --- & N & N & N & \\
7$_{3,4}$ -- 6$_{3,3}$ & 148709.018 & 55.4 & 19.1 & 17 & Y & --- & Y & N & N & \\
7$_{6,2}$ -- 6$_{6,1}$ & 149792.574 & 40.6 & 21.9 & 16 & N & --- & N & N & Y$^\ast$ & \\
7$_{1,6}$ -- 6$_{1,5}$ & 153432.176 & 32.5 & 25.1 & 16 & Y & --- & N & N & N & \\
13$_{2,11}$ -- 13$_{1,12}$ & 155894.300 & 105.9 & 1.26 & 16 & N & --- & N & N & N & \\
12$_{2,10}$ -- 12$_{1,11}$ & 155934.098 & 92.4 & 1.23 & 16 & N & --- & N & N & N & \\
11$_{2,9}$ -- 11$_{1,10}$ & 157072.457 & 79.9 & 1.22 & 16 & N & --- & N & N & N & \\
14$_{2,12}$ -- 14$_{1,13}$ & 157115.035 & 120.5 & 1.32 & 16 & N & --- & N & N & N & \\
10$_{2,8}$ -- 10$_{1,9}$ & 159127.569 & 68.4 & 1.21 & 15 & --- & --- & N & N & N & \\
15$_{2,13}$ -- 15$_{1,14}$ & 159739.080 & 136.2 & 1.39 & 15 & --- & --- & N & N & N & \\
9$_{2,7}$ -- 9$_{1,8}$ & 161899.774 & 58.1 & 1.22 & 15 & --- & --- & N & N & N & \\
8$_{1,8}$ -- 7$_{1,7}$ & 162958.657 & 38.2 & 30.5 & 15 & --- & --- & N & Y$^\ast$ & Y & \\
8$_{2,6}$ -- 8$_{1,7}$ & 165176.756 & 48.8 & 1.22 & 15 & --- & --- & N & N & N & \\
8$_{0,8}$ -- 7$_{0,7}$ & 167320.697 & 36.4 & 33.5 & 15 & --- & --- & N & Y$^\ast$ & Y & \\
7$_{2,5}$ -- 7$_{1,6}$ & 168741.408 & 40.6 & 1.23 & 15 & --- & --- & N & N & N & \\
8$_{2,7}$ -- 7$_{2,6}$ & 169299.154 & 48.5 & 32.6 & 15 & --- & --- & N & N & N & \\
8$_{6,2}$ -- 7$_{6,1}$ & 169790.683 & 143.9 & 15.3 & 14 & --- & --- & N & N & N & \\
8$_{6,3}$ -- 7$_{6,2}$ & 169790.683 & 143.9 & 15.3 & 14 & --- & --- & N & N & N & \\
8$_{5,4}$ -- 7$_{5,3}$ & 169810.709 & 111.1 & 21.4 & 14 & --- & --- & N & N & N & \\
8$_{5,3}$ -- 7$_{5,2}$ & 169810.715 & 111.1 & 21.4 & 14 & --- & --- & N & N & N & \\
8$_{4,5}$ -- 7$_{4,4}$ & 169861.469 & 84.3 & 26.3 & 14 & --- & --- & N & N & N & \\
8$_{4,4}$ -- 7$_{4,3}$ & 169862.523 & 84.3 & 26.3 & 14 & --- & --- & N & N & N & \\
8$_{3,6}$ -- 7$_{3,5}$ & 169955.835 & 63.5 & 30.2 & 14 & --- & --- & N & N & N & \\
8$_{3,5}$ -- 7$_{3,4}$ & 170039.076 & 63.5 & 30.3 & 14 & --- & --- & N & N & Y & \\
8$_{2,6}$ -- 7$_{2,5}$ & 171620.760 & 48.8 & 33.9 & 14 & --- & --- & N & N & N & \\
6$_{2,4}$ -- 6$_{1,5}$ & 172381.012 & 33.4 & 1.24 & 14 & --- & --- & N & N & N & \\
\hline
10$_{1,10}$ -- 9$_{1,9}$ & 203335.761 & 56.8 & 60.3 & 12 & Y & --- & N & N & Y$^\ast$ & \\
10$_{0,10}$ -- 9$_{0,9}$ & 207679.189 & 55.3 & 64.7 & 12 & Y & --- & N & Y$^\ast$ & N & \\
10$_{2,9}$ -- 9$_{2,8}$ & 211328.960 & 67.8 & 65.6 & 12 & Y & --- & Y & Y & Y$^\mathrm{b}$ & \\
10$_{5,6}$ -- 9$_{5,5}$ & 212323.555 & 130.5 & 52.0 & 12 & N & --- & N & N & N & \\
10$_{5,5}$ -- 9$_{5,4}$ & 212323.555 & 130.5 & 52.0 & 12 & N & --- & N & N & N & \\
10$_{4,7}$ -- 9$_{4,6}$ & 212428.020 & 103.7 & 58.4 & 12 & N & --- & N & N & B & NH$_2$CHO \\
10$_{4,6}$ -- 9$_{4,5}$ & 212433.449 & 103.7 & 58.4 & 12 & N & --- & N & N & B & NH$_2$CHO \\
10$_{3,8}$ -- 9$_{3,7}$ & 212572.837 & 82.9 & 63.3 & 12 & Y & --- & Y & N & N & \\
\hline
\end{tabular}
\label{tnh2cho}
\end{minipage}
\end{table*}

\begin{table*}
\centering
\begin{minipage}{175mm}
 \contcaption{}
\begin{tabular}{ccccccccccc}
\hline
Transition & $\nu$ & $E_u$ & $A_{ul}$ & $\theta_\mathrm{b}$ & OMC-2 & CepE & SVS13A & IRAS4A & I16293 & Blends\\
 & (MHz) & (K) & ($10^{-5}$~s$^{-1}$) & ($''$) &  &  &  &  &  & \\
\hline
10$_{3,7}$ -- 9$_{3,6}$ & 212832.307 & 82.9 & 63.6 & 12 & N & --- & N & N & N & \\
10$_{2,8}$ -- 9$_{2,7}$ & 215687.009 & 68.4 & 69.8 & 11 & Y & --- & N & N & Y & \\
10$_{1,9}$ -- 9$_{1,8}$ & 218459.213 & 60.8 & 74.7 & 11 & B & --- & N & N & N & CH$_3$OH\\
11$_{1,11}$ -- 10$_{1,10}$ & 223452.512 & 67.5 & 80.5 & 11 & Y & --- & N & N & N & \\
11$_{0,11}$ -- 10$_{0,10}$ & 227605.658 & 66.2 & 85.5 & 11 & Y & --- & N & N & N & \\
11$_{2,10}$ -- 10$_{2,9}$ & 232273.646 & 78.9 & 88.2 & 11 & N & --- & N & N & Y & \\
11$_{5,7}$ -- 10$_{5,6}$ & 233594.501 & 141.7 & 73.6 & 11 & N & --- & B & N & N & \\
11$_{5,6}$ -- 10$_{5,5}$ & 233594.501 & 141.7 & 73.6 & 11 & N & --- & B & N & N & \\
11$_{4,8}$ -- 10$_{4,7}$ & 233734.724 & 114.9 & 80.7 & 11 & N & --- & Y & N & N & \\
11$_{4,7}$ -- 10$_{4,6}$ & 233745.613 & 114.9 & 80.7 & 11 & N & --- & Y & N & N & \\
11$_{3,9}$ -- 10$_{3,8}$ & 233896.577 & 94.1 & 86.2 & 11 & Y & --- & Y & N & Y & \\
11$_{3,8}$ -- 10$_{3,7}$ & 234315.498 & 94.2 & 86.7 & 10 & --- & --- & N & N & N & \\
11$_{2,9}$ -- 10$_{2,8}$ & 237896.684 & 79.9 & 94.8 & 10 & Y & --- & Y & N & B & ? \\
11$_{1,10}$ -- 10$_{1,9}$ & 239951.800 & 72.3 & 99.6 & 10 & Y & --- & Y & N & N & \\
12$_{1,12}$ -- 11$_{1,11}$ & 243521.044 & 79.2 & 105 & 10 & N & --- & B & N & B & CH$_2$DOH \\
12$_{0,12}$ -- 11$_{0,11}$ & 247390.719 & 78.1 & 110 & 10 & N & --- & N & N & N & \\
12$_{2,11}$ -- 11$_{2,10}$ & 253165.793 & 91.1 & 115 & 10 & Y & --- & N & N & N & \\
12$_{4,9}$ -- 11$_{4,8}$ & 255058.533 & 127.2 & 108 & 10 & Y$^\ast$ & --- & Y$^\ast$ & N & N & \\
12$_{4,8}$ -- 11$_{4,7}$ & 255078.912 & 127.2 & 108 & 10 & N & --- & N & N & N & \\
12$_{3,10}$ -- 11$_{3,9}$ & 255225.651 & 106.4 & 114 & 10 & Y & --- & Y & N & N & \\
12$_{3,9}$ -- 11$_{3,8}$ & 255871.830 & 106.4 & 115 & 10 & Y & --- & N & N & N & \\
12$_{2,10}$ -- 11$_{2,9}$ & 260189.090 & 92.4 & 125 & 9 & B & --- & B & N & B & H$_2$C$_2$O\\
12$_{1,11}$ -- 11$_{1,10}$ & 261327.450 & 84.9 & 129 & 9 & N & --- & Y & N & N & \\
13$_{1,13}$ -- 12$_{1,12}$ & 263542.236 & 91.8 & 133 & 9 & Y & --- & Y & N & N & \\
\hline
\end{tabular}
\\
$^\mathrm{a}$Y: Detected above $T_\mathrm{mb} = 3\sigma$. Y$^\ast$: Weakly detected (S/N~$\sim 2-3$; see Sect.~\ref{lid}). N: undetected. B: possibly detected but blended. ---: not observed.\\
$^\mathrm{b}$Detected but with an anomalously high flux (maybe blended): removed from analysis.
\label{tnh2chob}
\end{minipage}
\end{table*}

\newpage

\begin{landscape}
\begin{table}
\centering
  \caption{HNCO transitions searched for in this study and 3$\sigma$ detections$^\mathrm{a}$}
\begin{tabular}{ccccccccccccccc}
\hline
Transition & $\nu$ & $E_u$ & $A_{ul}$ & $\theta_\mathrm{b}$ & OMC-2 & CepE & SVS13A & IRAS4A & I16293 & L1157 & L1527 & B1 & L1544 & TMC-1\\
 & (MHz) & (K) & ($10^{-5}$~s$^{-1}$) & ($''$) &  &  &  &  &  &  &  &  &  & \\
\hline
4$_{1,4}$ -- 3$_{1,3}$ & 87597.330 & 53.8 & 0.80 & 28 & N & N & Y & Y$^\mathrm{b}$ & N & N & N & N & N & ---\\
4$_{0,4}$ -- 3$_{0,3}$ & 87925.237 &10.5 & 0.88 & 28 & Y & Y & Y & Y & Y & Y & Y & Y & Y & ---\\
4$_{1,3}$ -- 3$_{1,2}$ & 88239.020 & 53.9 & 0.82 & 28 & Y$^\ast$ & N & Y$^\ast$ & Y$^\mathrm{b}$ & N & N & N & N & N & ---\\
5$_{1,5}$ -- 4$_{1,4}$ & 109495.996 & 59.0 & 1.7 & 22 & N & N & Y & N & Y & W & N & N & N & N\\
5$_{0,5}$ -- 4$_{0,4}$ & 109905.749 &15.8 & 1.8 & 22 & Y & Y & Y & Y & Y & Y & Y & Y & Y & Y\\
5$_{1,4}$ -- 4$_{1,3}$ & 110298.089 & 59.2 & 1.7 & 22 & Y$^\ast$ & N & Y & N & Y & N & N & N & N & N\\
6$_{1,6}$ -- 5$_{1,5}$ &  131394.230 & 65.3 & 2.9 & 19 & N & N & Y & N & Y & N & N & N & --- & N\\
6$_{0,6}$ -- 5$_{0,5}$ & 131885.734 & 22.2 & 3.1 & 19 & Y & Y & Y & Y & Y & Y & Y & Y & --- & Y\\
6$_{1,5}$ -- 5$_{1,4}$ & 132356.701 & 65.5 & 3.0 & 19 & N & N & N & N & Y & Y & N & N & --- & N\\
7$_{1,7}$ -- 6$_{1,6}$ & 153291.935 & 72.7 & 4.7 & 16 & N & --- & Y & N & Y & N & N & N & --- & N\\
7$_{0,7}$ -- 6$_{0,6}$ & 153865.086 & 29.5 & 4.9 & 16 & Y & --- & Y & Y & Y & Y & Y & Y & --- & Y\\
7$_{1,6}$ -- 6$_{1,5}$ & 154414.765 & 72.9 & 4.8 & 16 & N & --- & Y & N & Y & N & N & N & --- & N\\
10$_{1,10}$ -- 9$_{1,9}$ & 218981.009 & 101.1 & 14.2 & 11 & N & N & Y & N & Y & N & N & N & --- & ---\\
10$_{0,10}$ -- 9$_{0,9}$ & 219798.274 & 58.0 & 14.7 & 11 & Y & Y & Y & Y & Y & N & N & Y & --- & ---\\
10$_{1,9}$ -- 9$_{1,8}$ & 220584.751 & 101.5 & 14.5 & 11 & Y & N & Y & Y & Y & N & N & N & --- & ---\\
11$_{1,11}$ -- 10$_{1,10}$ & 240875.727 & 112.6 & 19.0 & 10 & N & --- & Y & Y & Y & N & N & N & --- & ---\\
11$_{0,11}$ -- 10$_{0,10}$ & 241774.032 & 69.6 & 19.6 & 10 & B & --- & B & B & B & N & N & N & --- & ---\\
11$_{1,10}$ -- 10$_{1,9}$ & 242639.704 & 113.1 & 19.5 & 10 & N & --- & Y & N & N & N & N & --- & ---\\
12$_{1,12}$ -- 11$_{1,11}$ & 262769.477 & 125.3 & 24.8 & 9 & N & N & Y & Y & Y$^\mathrm{c}$ & N & N & N & --- & ---\\
12$_{0,12}$ -- 11$_{0,11}$ & 263748.625 & 82.3 & 25.6 & 9 & Y & Y & Y & Y & Y & N & N & N & --- & ---\\
12$_{1,11}$ -- 11$_{1,10}$ & 264693.655 & 125.9 & 25.4 & 9 & N & N & Y & Y & Y & N & N & N & --- & ---\\
\hline
\end{tabular}
\\
$^\mathrm{a}$Y: Detected above $T_\mathrm{mb} = 3\sigma$. Y$^\ast$: Weakly detected (S/N~$\sim 2-3$; see Sect.~\ref{lid}). N: undetected. B: detected but blended. ---: not observed.\\
$^\mathrm{b}$Detected but with an anomalously high flux (maybe blended): removed from analysis.\\
$^\mathrm{c}$Blended with an unidentified feature: removed from analysis.\\
Blends: CH$_3$OH at 241.774~GHz\\
\label{thnco}
\end{table}
\end{landscape}

\newpage

\begin{table*}
\centering
\caption{L1544: Gaussian fits to the detected HNCO lines}
\begin{tabular}{ccccccc}
\hline
\hline
Transition & $\nu$ & RMS & $T_\mathrm{peak}$ & $V_\mathrm{lsr}$ & $\Delta V$ & $\int T_\mathrm{mb} dV$\\
 & (MHz) & (mK) & (mK) & (km~s$^{-1}$) & (km~s$^{-1}$) & (mK~km~s$^{-1}$) \\
\hline
4$_{0,4}$ -- 3$_{0,3}$ & 87925.237 & 11.9 & 459 (2) & 7.2 (1) & 0.7 (2) & 342 (9)\\
5$_{0,5}$ -- 4$_{0,4}$ & 109905.996 & 4.8 & 601 (2) & 7.6 (1) & 0.7 (1) & 448 (5)\\
\hline
\end{tabular}
\label{tl1544det}
\end{table*}

\begin{table*}
\centering
\caption{TMC-1: Gaussian fits to the detected HNCO lines}
\begin{tabular}{ccccccc}
\hline
\hline
Transition & $\nu$ & RMS & $T_\mathrm{peak}$ & $V_\mathrm{lsr}$ & $\Delta V$ & $\int T_\mathrm{mb} dV$\\
 & (MHz) & (mK) & (mK) & (km~s$^{-1}$) & (km~s$^{-1}$) & (mK~km~s$^{-1}$) \\
\hline
5$_{0,5}$ -- 4$_{0,4}$ & 109905.996 & 6.8 & 157 (8) & 5.8 (1) & 1.2 (1) & 203 (11)\\
6$_{0,6}$ -- 5$_{0,5}$ & 131885.734 & 6.7 & 94 (7) & 5.8 (1) & 0.5 (3) & 50 (5)\\
7$_{0,7}$ -- 6$_{0,6}$ & 153865.086 & 3.2 & 28 (3) & 5.9 (1) & 0.5 (4) & 15 (2)\\
\hline
\end{tabular}
\label{ttmc1det}
\end{table*}

\begin{table*}
\centering
\caption{B1: Gaussian fits to the detected HNCO lines}
\begin{tabular}{ccccccc}
\hline
\hline
Transition & $\nu$ & RMS & $T_\mathrm{peak}$ & $V_\mathrm{lsr}$ & $\Delta V$ & $\int T_\mathrm{mb} dV$\\
 & (MHz) & (mK) & (mK) & (km~s$^{-1}$) & (km~s$^{-1}$) & (mK~km~s$^{-1}$) \\
\hline
4$_{0,4}$ -- 3$_{0,3}$ & 87925.237 & 3.0 & 530 (12) & 6.7 (1) & 1.4 (1) & 765 (5)\\
5$_{0,5}$ -- 4$_{0,4}$ & 109905.996 & 22.2 & 480 (6) & 6.6 (1) & 1.3 (1) & 662 (32)\\
6$_{0,6}$ -- 5$_{0,5}$ & 131885.734 & 6.3 & 345 (7) & 6.7 (1) & 1.4 (1) & 521 (8)\\
7$_{0,7}$ -- 6$_{0,6}$ & 153865.086 & 7.4 & 224 (6) & 6.6 (1) & 1.4 (1) & 326 (10)\\
\hline
\end{tabular}
\label{tb1det}
\end{table*}

\begin{table*}
\centering
\caption{L1527: Gaussian fits to the detected HNCO lines}
\begin{tabular}{ccccccc}
\hline
\hline
Transition & $\nu$ & RMS & $T_\mathrm{peak}$ & $V_\mathrm{lsr}$ & $\Delta V$ & $\int T_\mathrm{mb} dV$\\
 & (MHz) & (mK) & (mK) & (km~s$^{-1}$) & (km~s$^{-1}$) & (mK~km~s$^{-1}$) \\
\hline
4$_{0,4}$ -- 3$_{0,3}$ & 87925.237 & 2.3 & 145 (2) & 5.9 (1) & 1.3 (1) & 198 (3)\\
5$_{0,5}$ -- 4$_{0,4}$ & 109905.996 & 7.9 & 135 (7) & 5.9 (1) & 1.2 (1) & 175 (11)\\
6$_{0,6}$ -- 5$_{0,5}$ & 131885.734 & 6.7 & 138 (1) & 5.9 (1) & 0.8 (1) & 115 (8)\\
7$_{0,7}$ -- 6$_{0,6}$ & 153865.086 & 8.0 & 63 (1) & 5.8 (1) & 0.7 (1) & 47 (6)\\
\hline
\end{tabular}
\label{tl1527det}
\end{table*}

\begin{table*}
\centering
\caption{L1157mm: Gaussian fits to the detected HNCO lines}
\begin{tabular}{ccccccc}
\hline
\hline
Transition & $\nu$ & RMS & $T_\mathrm{peak}$ & $V_\mathrm{lsr}$ & $\Delta V$ & $\int T_\mathrm{mb} dV$\\
 & (MHz) & (mK) & (mK) & (km~s$^{-1}$) & (km~s$^{-1}$) & (mK~km~s$^{-1}$) \\
\hline
4$_{0,4}$ -- 3$_{0,3}$ & 87925.237 & 3.3 & 113 (7) & 2.6 (1) & 1.6 (1) & 198 (7)\\
5$_{0,5}$ -- 4$_{0,4}$ & 109905.996 & 7.8 & 142 (5) & 2.5 (1) & 1.2 (1) & 177 (13)\\
6$_{0,6}$ -- 5$_{0,5}$ & 131885.734 & 5.5 & 81 (3) & 2.6 (1) & 1.5 (1) & 126 (11)\\
7$_{0,7}$ -- 6$_{0,6}$ & 153865.086 & 5.3 & 71 (1) & 2.6 (1) & 1.1 (1) & 81 (8)\\
\hline
\end{tabular}
\label{tl1157mmdet}
\end{table*}

\begin{table*}
\centering
\caption{IRAS~4A: Gaussian fits to the detected NH$_2$CHO and HNCO lines}
\begin{tabular}{ccccccc}
\hline
\hline
Transition & $\nu$ & RMS & $T_\mathrm{peak}$ & $V_\mathrm{lsr}$ & $\Delta V$ & $\int T_\mathrm{mb} dV$\\
 & (MHz) & (mK) & (mK) & (km~s$^{-1}$) & (km~s$^{-1}$) & (mK~km~s$^{-1}$) \\
\hline
\multicolumn{7}{c}{NH$_2$CHO}\\
\hline
5$_{0,5}$ -- 4$_{0,4}$ & 105464.219 & 3.3 & 9.8 (1) & 6.7 (4) & 4.0 (13) & 41 (11)\\
5$_{2,3}$ -- 4$_{2,2}$ & 106541.680 & 2.2 & 7.7 (17) & 7.6 (3) & 2.8 (9) & 23 (5) \\
7$_{1,7}$ -- 6$_{1,6}$ & 142701.325 & 5.0 & 15 (4) & 8.1 (3) & 2.6 (5) & 40 (8)\\
8$_{1,8}$ -- 7$_{1,7}$$^\mathrm{w}$ & 162958.657 & 38.2 & 14 (6) & 8.5 (4) & 2.9 (7) & 44 (9)\\
8$_{0,8}$ -- 7$_{0,7}$$^\mathrm{w}$ & 167320.697 & 36.4 & 13 (4) & 7.2 (4) & 3.2 (7) & 43 (10)\\
10$_{0,10}$ -- 9$_{0,9}$$^\mathrm{w}$ & 207679.189 & 55.3 & 19 (10) & 6.6 (5) & 3.3 (10) & 66 (19)\\
10$_{2,9}$ -- 9$_{2,8}$ & 211328.960 & 7.1 & 31 (3) & 6.7 (2) & 2.4 (5) & 82 (15)\\
\hline
\multicolumn{7}{c}{HNCO}\\
\hline
4$_{0,4}$ -- 3$_{0,3}$ & 87925.237 & 3.1 & 195 (16) & 7.2 (1) & 2.2 (1) & 458 (6)\\
5$_{0,5}$ -- 4$_{0,4}$ & 109905.749 & 8.0 & 198 (15) & 7.1 (1) & 2.3 (1) & 495 (16)\\
6$_{0,6}$ -- 5$_{0,5}$ & 131885.734 & 5.9 & 203 (12) & 7.0 (1) & 2.5 (1) & 545 (11)\\
7$_{0,7}$ -- 6$_{0,6}$ & 153865.086 & 6.8 & 168 (11) & 7.1 (1) & 2.2 (1) & 395 (12)\\
10$_{0,10}$ -- 9$_{0,9}$ & 219798.274 & 9.0 & 88 (8) & 6.8 (1) & 3.3 (3) & 307 (19)\\
10$_{1,9}$ -- 9$_{1,8}$ & 220584.751 & 6.8 & 37 (6) & 6.5 (3) & 2.8 (7) & 111 (22)\\
11$_{1,11}$ -- 10$_{1,10}$ & 240875.727 & 7.2 & 23 (2) & 6.6 (2) & 1.7 (6) & 42 (13)\\
12$_{1,12}$ -- 11$_{1,11}$ & 262769.477 & 8.9 & 44 (4) & 6.6 (2) & 3.8 (6) & 177 (21)\\
12$_{0,12}$ -- 11$_{0,11}$ & 263748.625 & 12.9 & 54 (4) & 6.5 (8) & 3.9 (29) & 230 (130)\\
12$_{1,11}$ -- 11$_{1,10}$ & 264693.655 & 9.1 & 26 (5) & 6.3 (4) & 3.7 (9) & 103 (21)\\
\hline
\end{tabular}
\\
$^\mathrm{w}$ Transition weakly detected (see Table~\ref{tnh2cho}) but included in the analysis for completeness.
\label{tiras4adet}
\end{table*}

\begin{table*}
\centering
\caption{I16293: Gaussian fits to the detected NH$_2$CHO and HNCO lines (intensity in $T_\mathrm{ant}^\ast$ units)}
\begin{tabular}{ccccccc}
\hline
\hline
Transition & $\nu$ & RMS & $T_\mathrm{peak}$ & $V_\mathrm{lsr}$ & $\Delta V$ & $\int T_\mathrm{a} dV$\\
 & (MHz) & (mK) & (mK) & (km~s$^{-1}$) & (km~s$^{-1}$) & (mK~km~s$^{-1}$) \\
\hline
\multicolumn{7}{c}{NH$_2$CHO}\\
\hline
4$_{0,4}$ -- 3$_{0,3}$ & 84542.330 & 5.3 & 17 (6) & 3.2 (12) & 6.7 (20) & 120 (40)\\
4$_{1,3}$ -- 3$_{1,2}$ & 87848.873 & 2.6 & 12 (4) & 2.6 (5) & 3.8 (11) & 50 (13)\\
5$_{1,5}$ -- 4$_{1,4}$ & 102064.267 & 3.3 & 16 (5) & 2.0 (3) & 2.4 (9) & 42 (12)\\
5$_{0,5}$ -- 4$_{0,4}$ & 105464.219 & 5.3 & 20 (7) & 5.1 (4) & 3.0 (10) & 64 (19)\\
7$_{6,2}$ -- 6$_{6,1}$$^\mathrm{w}$ & 149792.574 & 10.0 & 29 (10) & 2.6 (4) & 3.0 (7) & 93 (22)\\
8$_{1,8}$ -- 7$_{1,7}$ & 162958.657 & 10.1 & 26 (10) & 2.8 (4) & 2.5 (13) & 69 (27)\\
8$_{0,8}$ -- 7$_{0,7}$ & 167320.697 & 10.6 & 55 (11) & 2.0 (2) & 2.3 (5) & 136 (24)\\
8$_{3,5}$ -- 7$_{3,4}$ & 170039.076 & 15.3 & 50 (80) & 2.7 (17) & 2.4 (46) & 120 (170)\\
10$_{1,10}$ -- 9$_{1,9}$$^\mathrm{w}$ & 203335.761 & 6.8 & 19 (7) & 0.5 (10) & 6.7 (27) & 135 (43)\\
10$_{2,8}$ -- 9$_{2,7}$ & 215687.009 & 5.5 & 30 (10) & 1.0 (13) & 5.5 (34) & 175 (90)\\
11$_{2,10}$ -- 10$_{2,9}$ & 232273.646 & 4.6 & 23 (18) & 1.5 (19) & 5.0 (39) & 120 (90)\\
11$_{2,9}$ -- 10$_{2,8}$ & 237896.684 & 8.2 & 45 (9) & 2.8 (3) & 3.1 (15) & 148 (42)\\
\hline
\multicolumn{7}{c}{HNCO}\\
\hline
4$_{0,4}$ -- 3$_{0,3}$ & 87925.237 & 3.2 & 162 (3) & 4.0 (1) & 3.9 (1) & 671 (11)\\
5$_{1,5}$ -- 4$_{1,4}$ & 109495.996 & 5.4 & 38 (6) & 2.0 (3) & 4.5 (11) & 182 (35)\\
5$_{0,5}$ -- 4$_{0,4}$ & 109905.749 & 5.9 & 254 (6) & 3.9 (1) & 3.2 (1) & 853 (18)\\
5$_{1,4}$ -- 4$_{1,3}$ & 110298.089 & 5.4 & 16 (5) & 5.4 (6) & 5.4 (18) & 90 (23)\\
6$_{1,6}$ -- 5$_{1,5}$ &  131394.230 & 5.4 & 39 (17) & 2.5 (8) & 5.2 (27) & 217 (88)\\
6$_{0,6}$ -- 5$_{0,5}$ & 131885.734 & 7.3 & 263 (8) & 3.9 (1) & 3.0 (1) & 825 (18)\\
6$_{1,5}$ -- 5$_{1,4}$ & 132356.701 & 6.9 & 39 (8) & 3.6 (5) & 5.3 (13) & 218 (48)\\
7$_{1,7}$ -- 6$_{1,6}$ & 153291.935 & 9.5 & 73 (27) & 5.1 (5) & 3.0 (16) & 230 (90)\\
7$_{0,7}$ -- 6$_{0,6}$ & 153865.086 & 12.3 & 224 (18) & 3.9 (1) & 4.0 (3) & 950 (50)\\
7$_{1,6}$ -- 6$_{1,5}$ & 154414.765 & 11.1 & 57 (28) & 2.7 (7) & 3.4 (21) & 200 (90)\\
10$_{1,10}$ -- 9$_{1,9}$ & 218981.009 & 6.2 & 103 (8) & 2.8 (2) & 6.0 (5) & 664 (47)\\
10$_{0,10}$ -- 9$_{0,9}$ & 219798.274 & 4.4 & 260 (12) & 3.3 (1) & 5.7 (3) & 1580 (60)\\
10$_{1,9}$ -- 9$_{1,8}$ & 220584.751 & 6.4 & 85 (10) & 2.9 (3) & 6.2 (7) & 560 (50)\\
11$_{1,11}$ -- 10$_{1,10}$ & 240875.727 & 14.5 & 132 (16) & 2.8 (3) & 6.6 (9) & 930 (100)\\
12$_{0,12}$ -- 11$_{0,11}$ & 263748.625 & 7.9 & 270 (70) & 3.7 (6) & 5.6 (15) & 1640 (360)\\
12$_{1,11}$ -- 11$_{1,10}$ & 264693.655 & 6.9 & 119 (11) & 3.4 (2) & 6.6 (6) & 840 (60)\\
\hline
\end{tabular}
\\
$^\mathrm{w}$ Transition weakly detected (see Table~\ref{tnh2cho}) but included in the analysis for completeness.
\label{ti16293det}
\end{table*}

\begin{table*}
\centering
\caption{SVS13A: Gaussian fits to the detected NH$_2$CHO and HNCO lines}
\begin{tabular}{ccccccc}
\hline
\hline
Transition & $\nu$ & RMS & $T_\mathrm{peak}$ & $V_\mathrm{lsr}$ & $\Delta V$ & $\int T_\mathrm{mb} dV$\\
 & (MHz) & (mK) & (mK) & (km~s$^{-1}$) & (km~s$^{-1}$) & (mK~km~s$^{-1}$) \\
\hline
\multicolumn{7}{c}{NH$_2$CHO}\\
\hline
5$_{0,5}$ -- 4$_{0,4}$ & 105464.219 & 3.9 & 16 (1) & 8.1 (2) & 1.8 (5) & 14 (1)\\
7$_{1,7}$ -- 6$_{1,6}$ & 142701.325 & 6.4 & 28 (3) & 8.6 (2) & 2.2 (5) & 22 (5)\\
7$_{3,4}$ -- 6$_{3,3}$ & 148709.018 & 7.1 & 26 (2) & 7.1 (3) & 2.2 (7) & 22 (3)\\
10$_{2,9}$ -- 9$_{2,8}$ & 211328.960 & 7.3 & 62 (5) & 7.5 (5) & 4.4 (14) & 31 (2)\\
10$_{3,8}$ -- 9$_{3,7}$ & 212572.837 & 7.2 & 43 (7) & 8.4 (2) & 2.7 (6) & 29 (2)\\
11$_{4,8}$ -- 10$_{4,7}$ & 233734.724 & 6.6 & 26 (2) & 6.9 (4) & 3.0 (9) & 23 (5)\\
11$_{4,7}$ -- 10$_{4,6}$ & 233745.613 & 7.5 & 26 (2) & 8.2 (3) & 1.5 (7) & 26 (1)\\
11$_{3,9}$ -- 10$_{3,8}$ & 233896.577 & 6.6 & 49 (14) & 8.2 (4) & 2.9 (9) & 20 (3)\\
11$_{2,9}$ -- 10$_{2,8}$ & 237896.684 & 6.9 & 48 (6) & 8.1 (3) & 4.3 (8) & 28 (4)\\
11$_{1,10}$ -- 10$_{1,9}$ & 239951.800 & 8.1 & 66 (7) & 7.6 (4) & 3.4 (9) & 23 (2)\\
12$_{3,10}$ -- 11$_{3,9}$ & 255225.651 & 5.6 & 45 (2) & 8.8 (12) & 3.3 (36) & 24 (4)\\
12$_{1,11}$ -- 11$_{1,10}$ & 261327.450 & 8.7 & 26 (2) & 8.4 (2) & 4.6 (6) & 38 (5)\\
13$_{1,13}$ -- 12$_{1,12}$ & 263542.236 & 7.6 & 54 (5) & 7.7 (6) & 4.4 (15) & 32 (5)\\
\hline
\multicolumn{7}{c}{HNCO}\\
\hline
4$_{1,4}$ -- 3$_{1,3}$ & 87597.330 & 3.5 & 11 (3) & 6.4 (4) & 3.3 (8) & 40 (9)\\
4$_{0,4}$ -- 3$_{0,3}$ & 87925.237 & 9.1 & 46 (4) & 8.6 (2) & 3.1 (6) & 155 (23)\\
4$_{1,3}$ -- 3$_{1,2}$$^\mathrm{w}$ & 88239.020 & 3.0 & 13 (3) & 7.3 (4) & 3.3 (20) & 45 (17)\\
5$_{1,5}$ -- 4$_{1,4}$ & 109495.996 & 3.9 & 16 (1) & 7.1 (2) & 2.0 (5) & 35 (8)\\
5$_{0,5}$ -- 4$_{0,4}$$^\ast$ & 109905.749 & 5.4 & 76 (5) & 8.5 (1) & 1.2 (1) & 100 (8)\\
5$_{1,4}$ -- 4$_{1,3}$ & 110298.089 & 6.2 & 28 (4) & 8.2 (3) & 4.7 (8) & 139 (19)\\
6$_{1,6}$ -- 5$_{1,5}$ &  131394.230 & 6.6 & 19 (3) & 8.2 (4) & 3.6 (8) & 72 (15)\\
6$_{0,6}$ -- 5$_{0,5}$$^\ast$ & 131885.734 & 6.0 & 77 (10) & 8.5 (1) & 1.6 (3) & 129 (15)\\
7$_{1,7}$ -- 6$_{1,6}$ & 153291.935 & 5.7 & 25 (2) & 8.5 (5) & 5.1 (13) & 140 (30)\\
7$_{0,7}$ -- 6$_{0,6}$$^\ast$ & 153865.086 & 6.1 & 70 (6) & 7.5 (1) & 3.4 (5) & 258 (19)\\
7$_{1,6}$ -- 6$_{1,5}$ & 154414.765 & 6.0 & 32 (3) & 8.3 (6) & 4.4 (15) & 149 (40)\\
10$_{1,10}$ -- 9$_{1,9}$ & 218981.009 & 6.8 & 46 (3) & 8.2 (2) & 4.3 (4) & 212 (15)\\
10$_{0,10}$ -- 9$_{0,9}$ & 219798.274 & 7.1 & 89 (4) & 8.4 (1) & 3.2 (2) & 309 (16)\\
10$_{1,9}$ -- 9$_{1,8}$ & 220584.751 & 6.0 & 35 (3) & 8.5 (4) & 3.7 (9) & 136 (27)\\
11$_{1,11}$ -- 10$_{1,10}$ & 240875.727 & 6.7 & 46 (4) & 8.2 (5) & 4.9 (17) & 236 (66)\\
11$_{1,10}$ -- 10$_{1,9}$ & 242639.704 & 8.7 & 42 (5) & 8.1 (3) & 4.6 (6) & 206 (24)\\
12$_{1,12}$ -- 11$_{1,11}$ & 262769.477 & 9.9 & 68 (4) & 8.5 (2) & 5.3 (5) & 380 (32)\\
12$_{0,12}$ -- 11$_{0,11}$ & 263748.625 & 9.3 & 60 (5) & 8.2 (4) & 3.4 (9) & 213 (48)\\
12$_{1,11}$ -- 11$_{1,10}$ & 264693.655 & 8.9 & 39 (4) & 8.5 (3) & 4.4 (6) & 183 (23)\\
\hline
\end{tabular}
\\
$^\ast$ Transition affected by emission at OFF position: lower limit point in the rotational diagram.\\
$^\mathrm{w}$ Transition weakly detected (see Table~\ref{tnh2cho}) but included in the analysis for completeness.
\label{tsvsdet}
\end{table*}

\begin{table*}
\centering
\caption{Cep~E: Gaussian fits to the detected NH$_2$CHO and HNCO lines}
\begin{tabular}{ccccccc}
\hline
\hline
Transition & $\nu$ & RMS & $T_\mathrm{peak}$ & $V_\mathrm{lsr}$ & $\Delta V$ & $\int T_\mathrm{mb} dV$\\
 & (MHz) & (mK) & (mK) & (km~s$^{-1}$) & (km~s$^{-1}$) & (mK~km~s$^{-1}$) \\
\hline
\multicolumn{7}{c}{NH$_2$CHO}\\
\hline
4$_{0,4}$ -- 3$_{0,3}$ & 84542.330 & 1.9 & 7.7 (1) & --11.5 (4) & 2.8 (10) & 23 (7)\\
4$_{2,2}$ -- 3$_{2,1}$ & 85093.272 & 1.3 & 4.3 (1) & --11.8 (3) & 3.5 (7) & 16 (3)\\
4$_{1,3}$ -- 3$_{1,2}$$^\mathrm{w}$ & 87848.873 & 1.2 & 6.3 (1) & --10.5 (2) & 2.6 (5) & 17 (3)\\
5$_{1,5}$ -- 4$_{1,4}$ & 102064.267 & 1.6 & 7.6 (1) & --10.6 (2) & 1.7 (5) & 13 (3)\\
5$_{0,5}$ -- 4$_{0,4}$ & 105464.219 & 2.1 & 4.1 (1) & --10.6 (6) & 2.9 (11) & 13 (4)\\
\hline
\multicolumn{7}{c}{HNCO}\\
\hline
4$_{0,4}$ -- 3$_{0,3}$ & 87925.237 & 1.4 & 90 (4) & --11.1 (1) & 1.9 (1) & 179 (3)\\
5$_{0,5}$ -- 4$_{0,4}$ & 109905.749 & 2.8 & 104 (12) & --11.1 (1) & 2.4 (1) & 262 (6)\\
6$_{0,6}$ -- 5$_{0,5}$ & 131885.734 & 7.8 & 128 (9) & --11.1 (1) & 2.2 (3) & 299 (23)\\
10$_{0,10}$ -- 9$_{0,9}$ & 219798.274 & 7.5 & 45 (10) & --10.1 (73) & 8.0 (80) & 380 (65)\\
12$_{0,12}$ -- 11$_{0,11}$ & 263748.625 & 6.0 & 38 (8) & --10.2 (5) & 5.6 (14) & 226 (45)\\
\hline
\end{tabular}
\\
$^\mathrm{w}$ Transition weakly detected (see Table~\ref{tnh2cho}) but included in the analysis for completeness.
\label{tcepedet}
\end{table*}

\begin{table*}
\centering
\caption{OMC-2~FIR~4: Gaussian fits to the detected NH$_2$CHO and HNCO lines}
\begin{tabular}{ccccccc}
\hline
\hline
Transition & $\nu$ & RMS & $T_\mathrm{peak}$ & $V_\mathrm{lsr}$ & $\Delta V$ & $\int T_\mathrm{mb} dV$\\
 & (MHz) & (mK) & (mK) & (km~s$^{-1}$) & (km~s$^{-1}$) & (mK~km~s$^{-1}$) \\
\hline
\multicolumn{7}{c}{NH$_2$CHO}\\
\hline
4$_{0,4}$ -- 3$_{0,3}$ & 84542.330 & 2.8 & 15 (2) & 10.8 (2) & 2.2 (4) & 35 (5)\\
4$_{1,3}$ -- 3$_{1,2}$ & 87848.873 & 2.8 & 9 (2) & 11.0 (4) & 3.6 (6) & 37 (7)\\
5$_{1,5}$ -- 4$_{1,4}$ & 102064.267 & 3.5 & 15 (1) & 11.0 (2) & 2.9 (5) & 46 (6)\\
5$_{0,5}$ -- 4$_{0,4}$ & 105464.219 & 4.7 & 16 (1) & 11.5 (4) & 3.7 (10) & 64 (15)\\
5$_{2,4}$ -- 4$_{2,3}$ & 105972.599 & 4.6 & 15 (5) & 11.3 (2) & 1.7 (5) & 28 (8)\\
6$_{1,5}$ -- 5$_{1,4}$ & 131617.902 & 5.6 & 29 (3) & 11.2 (2) & 3.8 (6) & 115 (13)\\
7$_{1,7}$ -- 6$_{1,6}$ & 142701.325 & 6.4 & 28 (3) & 11.5 (2) & 3.1 (7) & 91 (14)\\
7$_{3,4}$ -- 6$_{3,3}$ & 148709.018 & 5.6 & 26 (2) & 11.5 (2) & 1.8 (5) & 51 (10)\\
7$_{1,6}$ -- 6$_{1,5}$ & 153432.176 & 8.2 & 40 (3) & 11.8 (3) & 2.7 (10) & 114 (31)\\
10$_{1,10}$ -- 9$_{1,9}$ & 203335.761 & 10.8 & 52 (4) & 11.3 (2) & 2.8 (5) & 156 (20)\\
10$_{0,10}$ -- 9$_{0,9}$ & 207679.189 & 8.8 & 49 (3) & 11.6 (1) & 1.8 (4) & 97 (14)\\
10$_{3,8}$ -- 9$_{3,7}$ & 212572.837 & 12.7 & 43 (7) & 11.7 (3) & 2.9 (6) & 131 (23)\\
11$_{1,11}$ -- 10$_{1,10}$ & 223452.512 & 12.6 & 44 (6) & 11.7 (5) & 3.0 (12) & 142 (48)\\
11$_{0,11}$ -- 10$_{0,10}$ & 227605.658 & 14.5 & 61 (10) & 11.9 (3) & 3.7 (7) & 237 (33)\\
11$_{3,9}$ -- 10$_{3,8}$ & 233896.577 & 16.9 & 49 (14) & 11.6 (3) & 2.9 (10) & 151 (36)\\
11$_{2,9}$ -- 10$_{2,8}$ & 237896.684 & 10.1 & 48 (6) & 11.4 (2) & 3.7 (6) & 186 (23)\\
11$_{1,10}$ -- 10$_{1,9}$ & 239951.800 & 10.5 & 66 (7) & 11.5 (1) & 2.5 (4) & 175 (21)\\
12$_{2,11}$ -- 11$_{2,10}$ & 253165.793 & 12.4 & 44 (9) & 11.2 (3) & 3.2 (7) & 152 (27)\\
12$_{4,9}$ -- 11$_{4,8}$$^\mathrm{w}$ & 255058.533 & 12.5 & 39 (12) & 11.6 (2) & 1.9 (4) & 80 (17)\\
12$_{3,10}$ -- 11$_{3,9}$ & 255225.651 & 11.4 & 45 (2) & 11.5 (4) & 1.6 (9) & 74 (4)\\
12$_{3,9}$ -- 11$_{3,8}$ & 255871.830 & 11.8 & 41 (7) & 11.8 (2) & 2.7 (5) & 119 (19)\\
\hline
\multicolumn{7}{c}{HNCO}\\
\hline
4$_{0,4}$ -- 3$_{0,3}$ & 87925.237 & 2.8 & 128 (13) & 11.2 (1) & 2.5 (1) & 344 (6)\\
4$_{1,3}$ -- 3$_{1,2}$$^\mathrm{w}$ & 88239.020 & 2.3 & 9 (2) & 12.7 (3) & 3.1 (6) & 31 (5)\\
5$_{0,5}$ -- 4$_{0,4}$ & 109905.749 & 6.9 & 224 (22) & 11.4 (1) & 2.1 (1) & 512 (14)\\
5$_{1,4}$ -- 4$_{1,3}$$^\mathrm{w}$ & 110298.089 & 6.2 & 18 (5) & 12.8 (2) & 1.4 (5) & 26 (7)\\
6$_{0,6}$ -- 5$_{0,5}$ & 131885.734 & 4.7 & 267 (27) & 11.2 (1) & 2.7 (1) & 775 (11)\\
7$_{0,7}$ -- 6$_{0,6}$ & 153865.086 & 8.8 & 304 (38) & 11.2 (1) & 3.0 (4) & 962 (89)\\
10$_{0,10}$ -- 9$_{0,9}$ & 219798.274 & 9.2 & 343 (28) & 11.4 (1) & 2.5 (1) & 918 (17)\\
10$_{1,9}$ -- 9$_{1,8}$ & 220584.751 & 11.0 & 36 (1) & 11.7 (4) & 2.1 (13) & 79 (35)\\
12$_{0,12}$ -- 11$_{0,11}$ & 263748.625 & 5.1 & 202 (9) & 11.9 (3) & 3.0 (5) & 648 (24)\\
\hline
\end{tabular}
\\
$^\mathrm{w}$ Transition weakly detected (see Table~\ref{tnh2cho}) but included in the analysis for completeness.
\label{tomc2det}
\end{table*}

\clearpage

\section[]{Figures}

\begin{figure*}
\centering
\begin{tabular}{lr}
   & \includegraphics[scale=0.55]{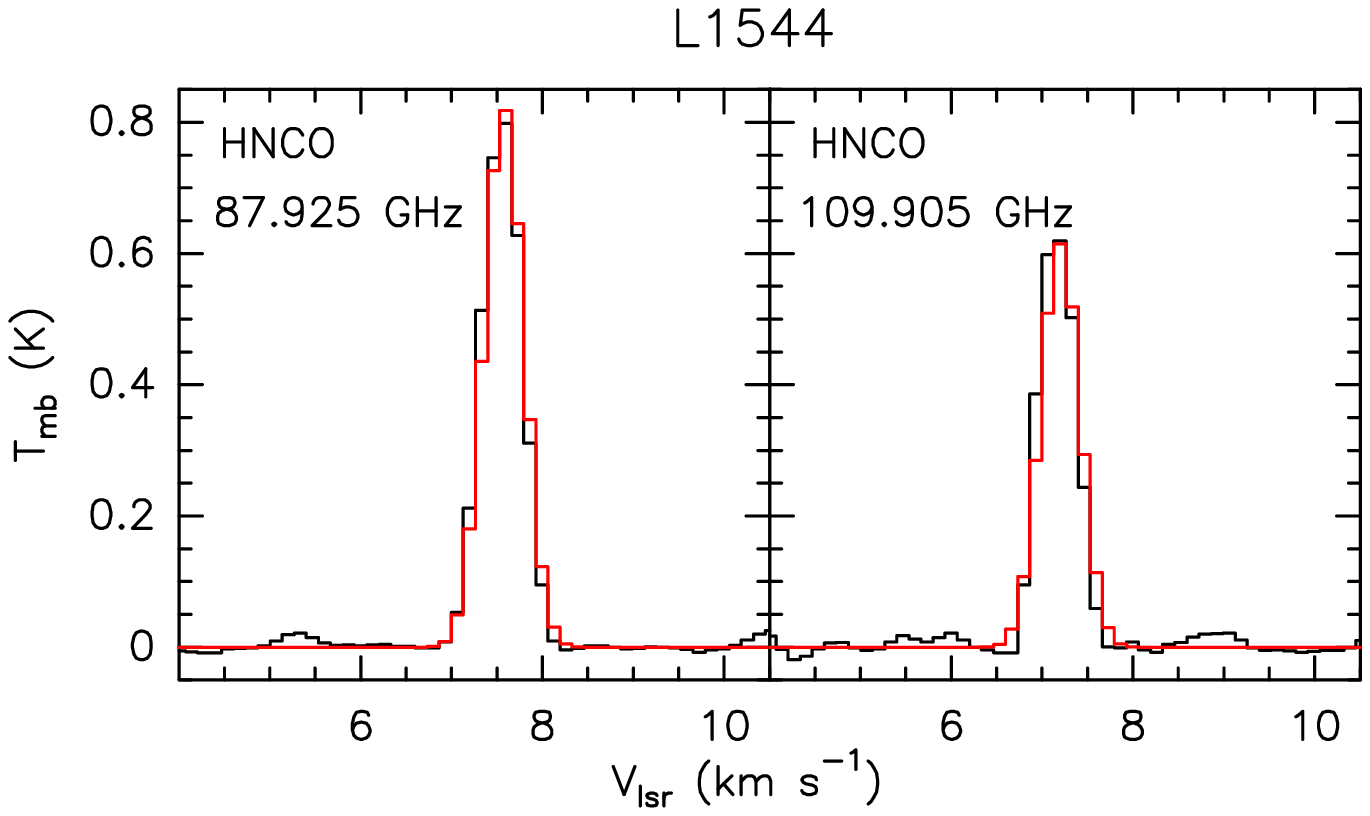}\\
 \includegraphics[scale=0.55]{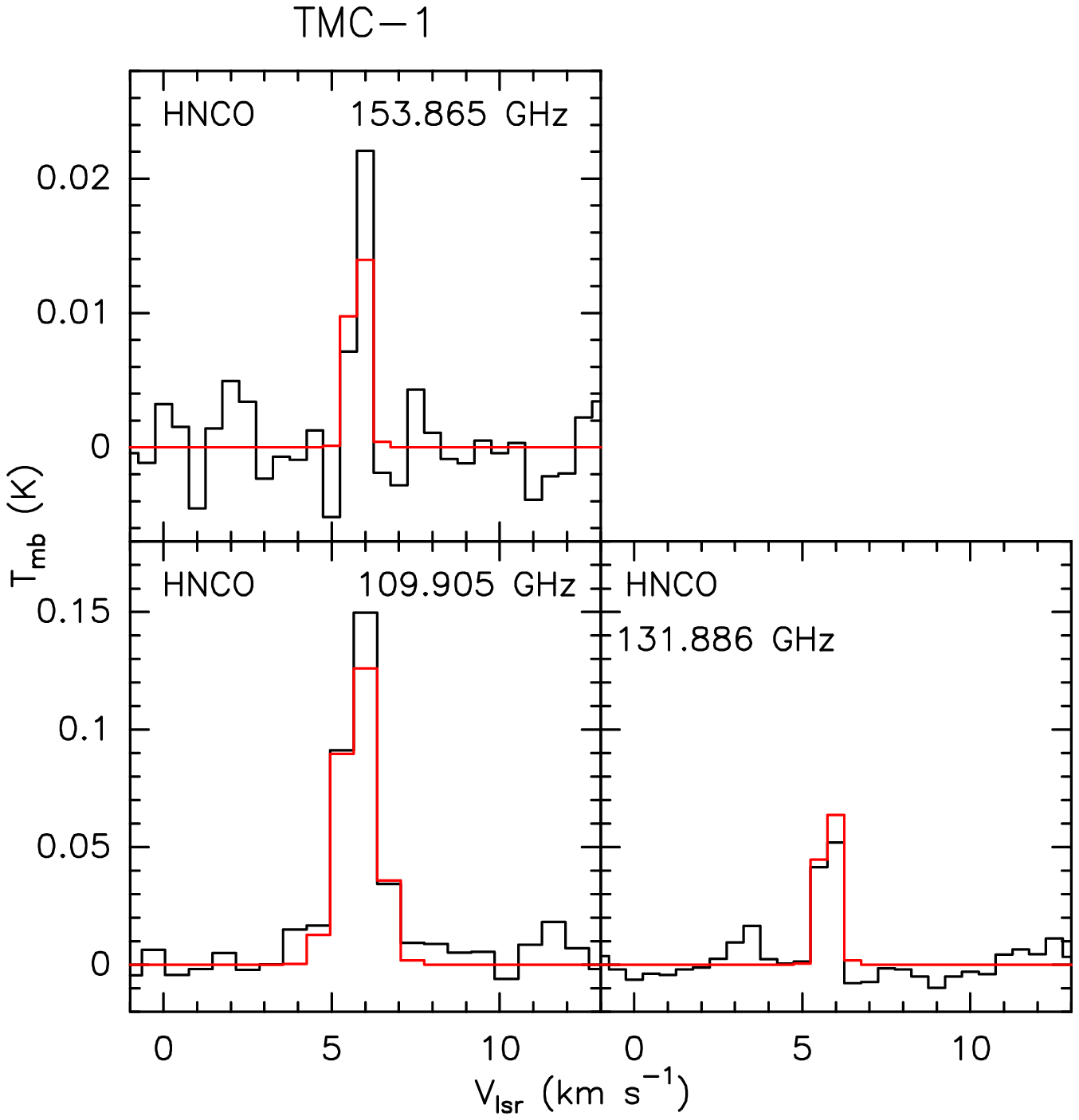} & \includegraphics[scale=0.55]{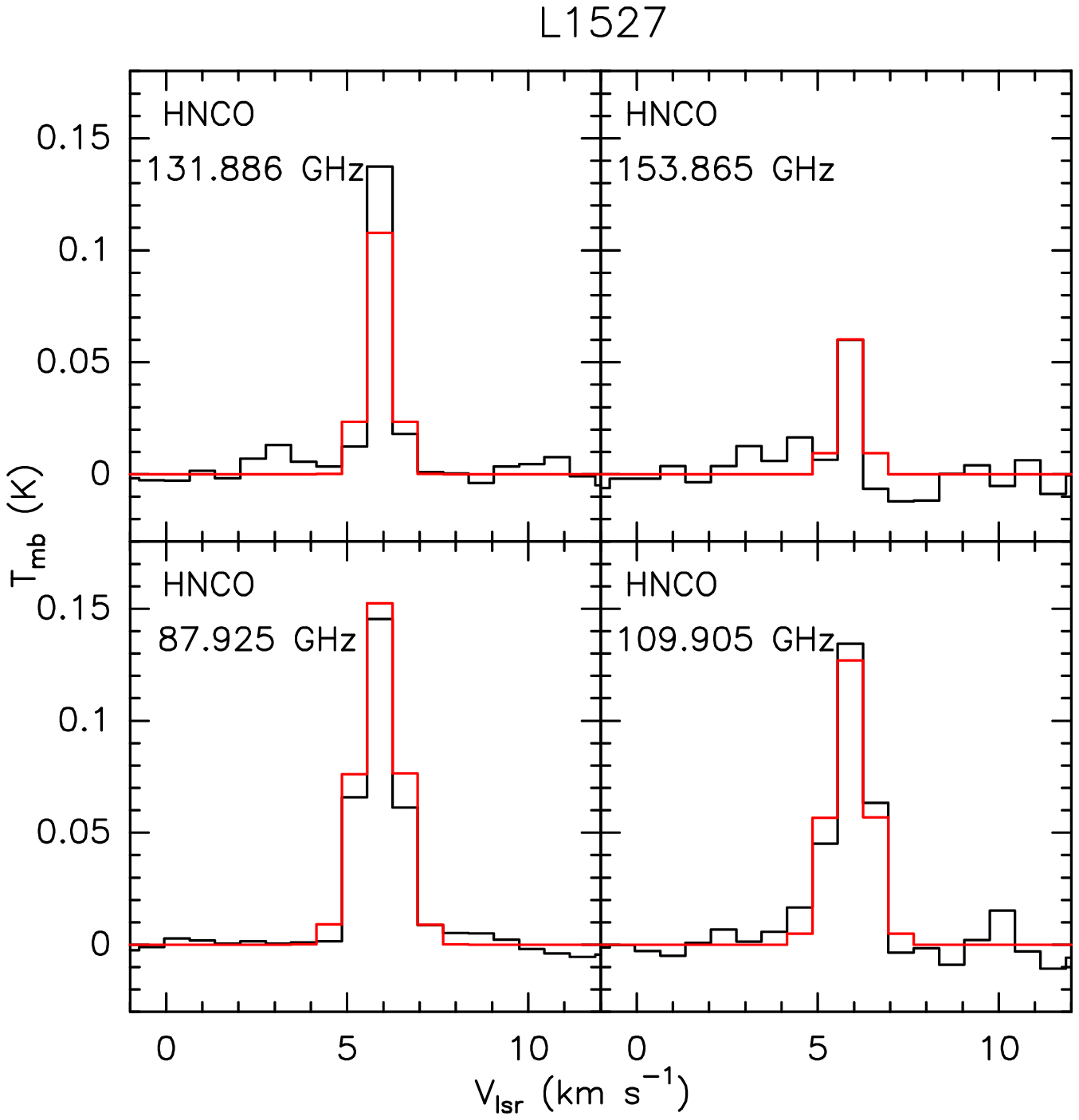}\\
   \includegraphics[scale=0.55]{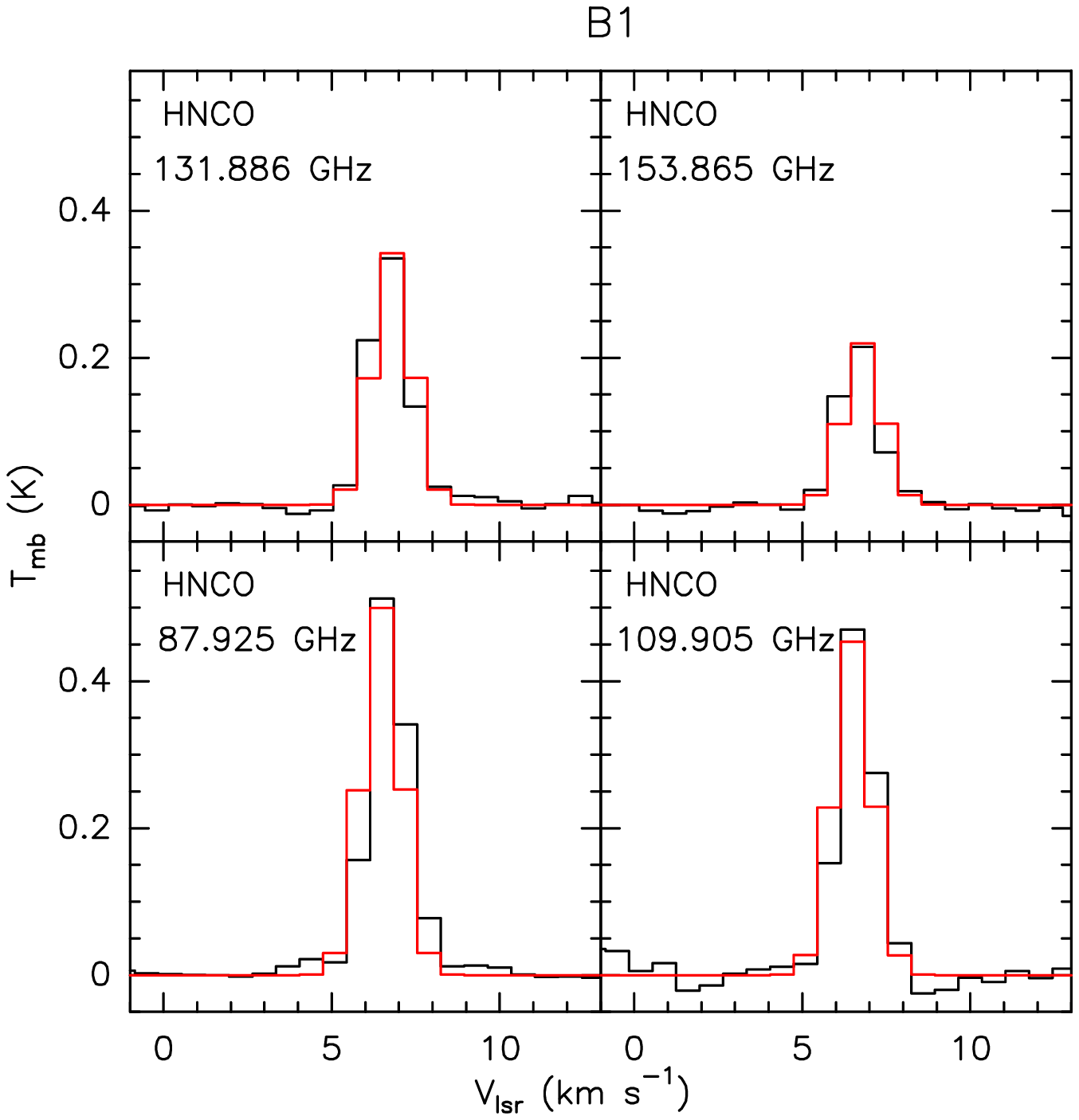} & \includegraphics[scale=0.55]{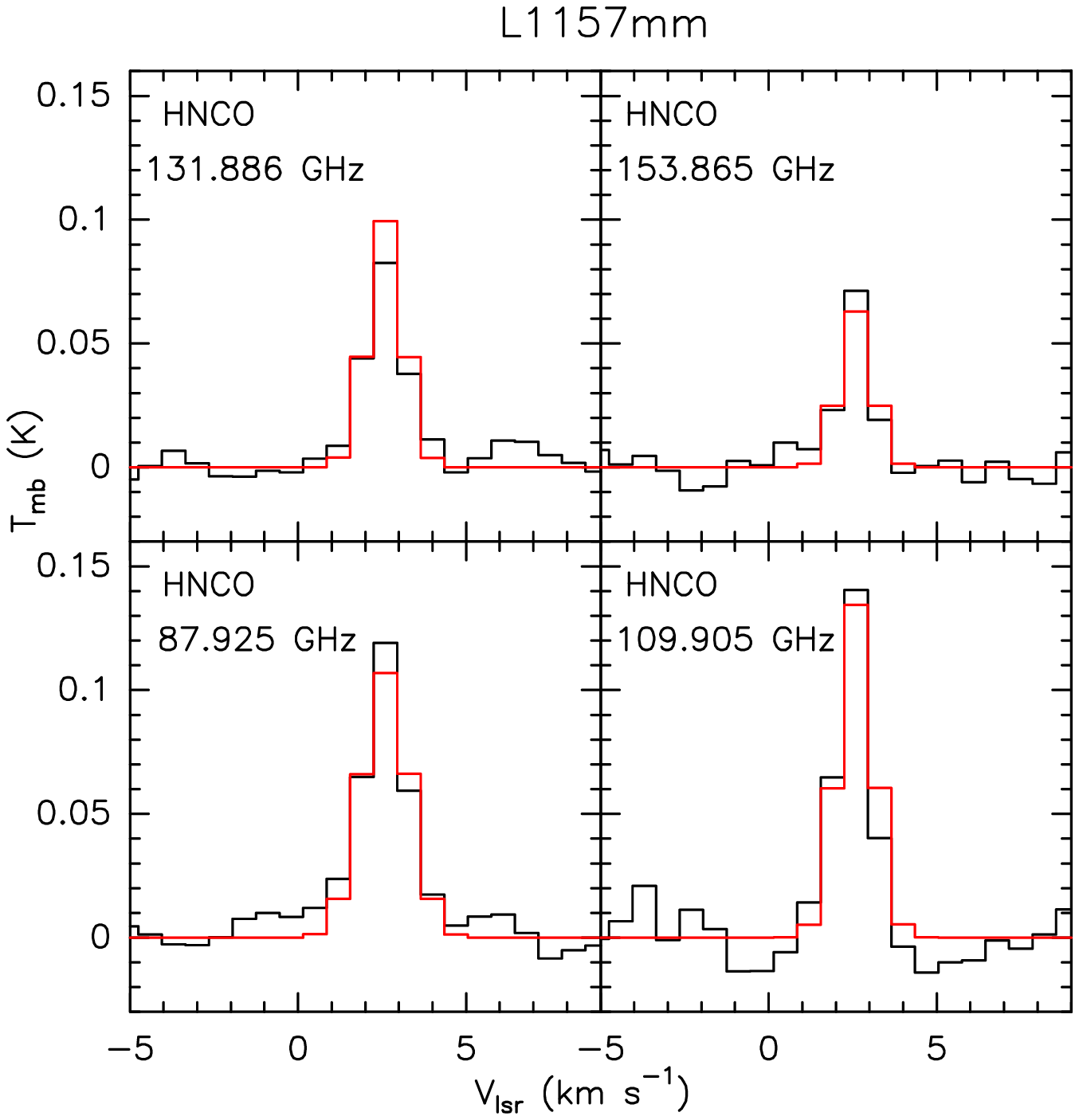}\\
\end{tabular}
  \caption{HNCO observed spectral lines (black) in L1544, TMC-1, B1, L1527, and L1157mm, and the spectra predicted by best fit LTE model (red).}
  \label{fspt1}
\end{figure*}

\newpage

\begin{figure*}
\centering
\begin{tabular}{lr}
   \includegraphics[scale=0.55]{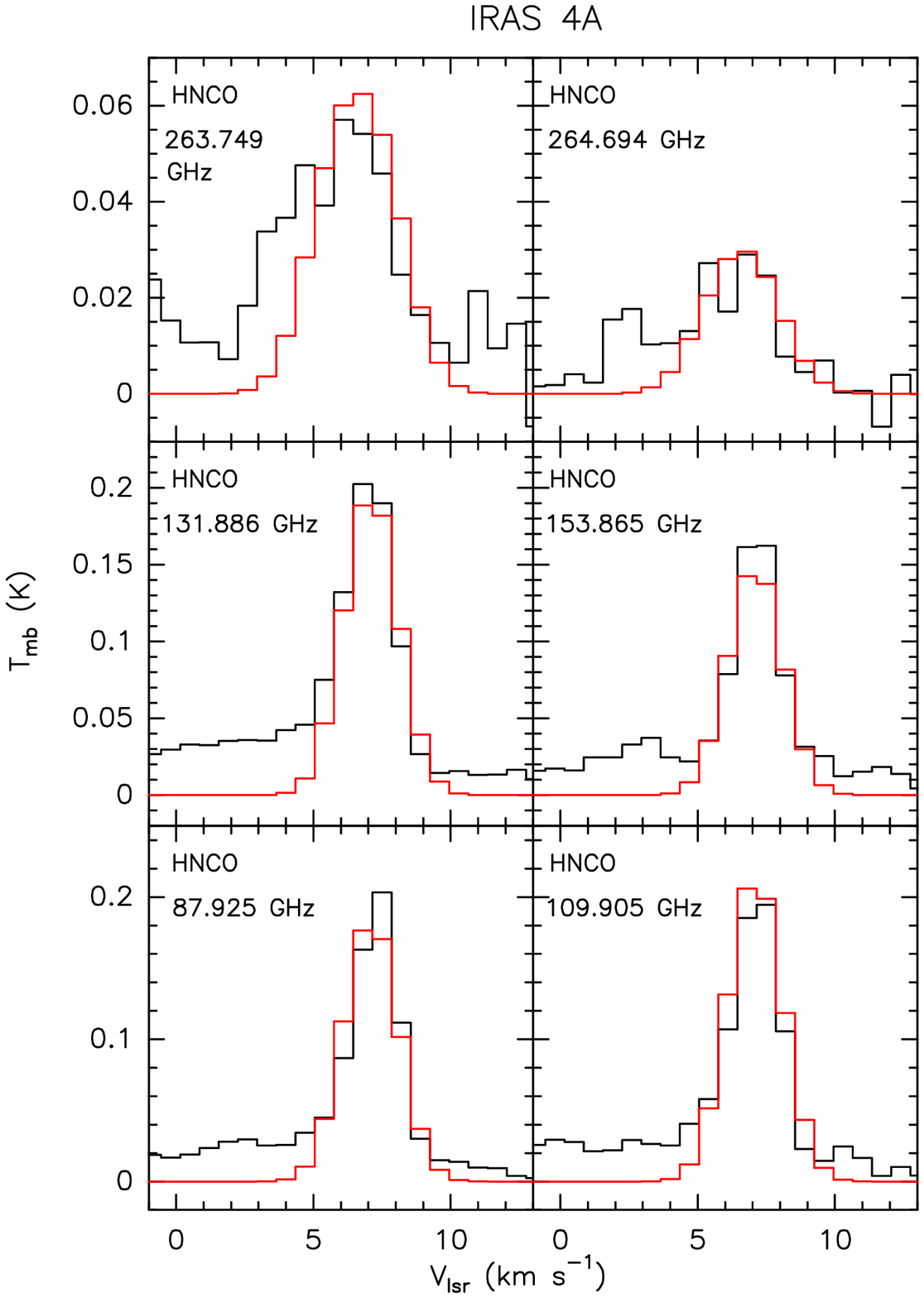} & \includegraphics[scale=0.55]{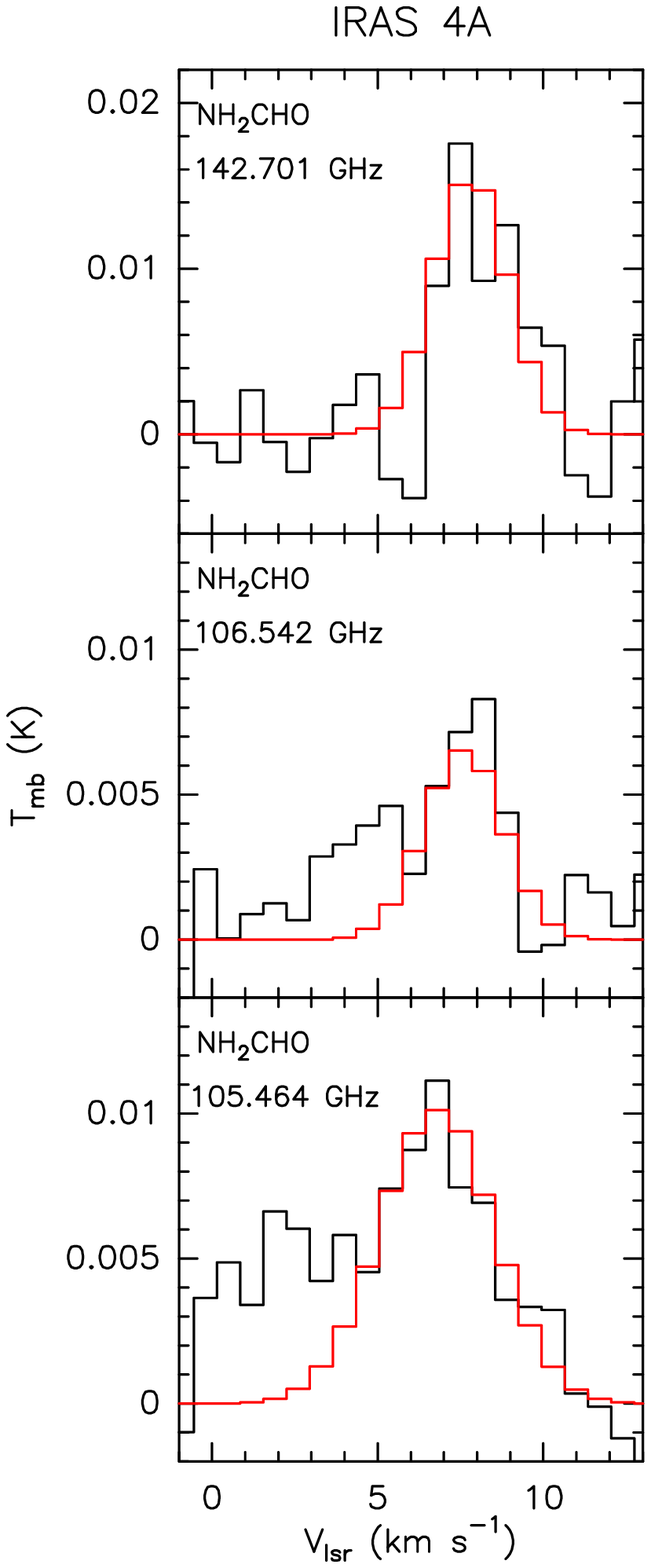}\\
   \includegraphics[scale=0.55]{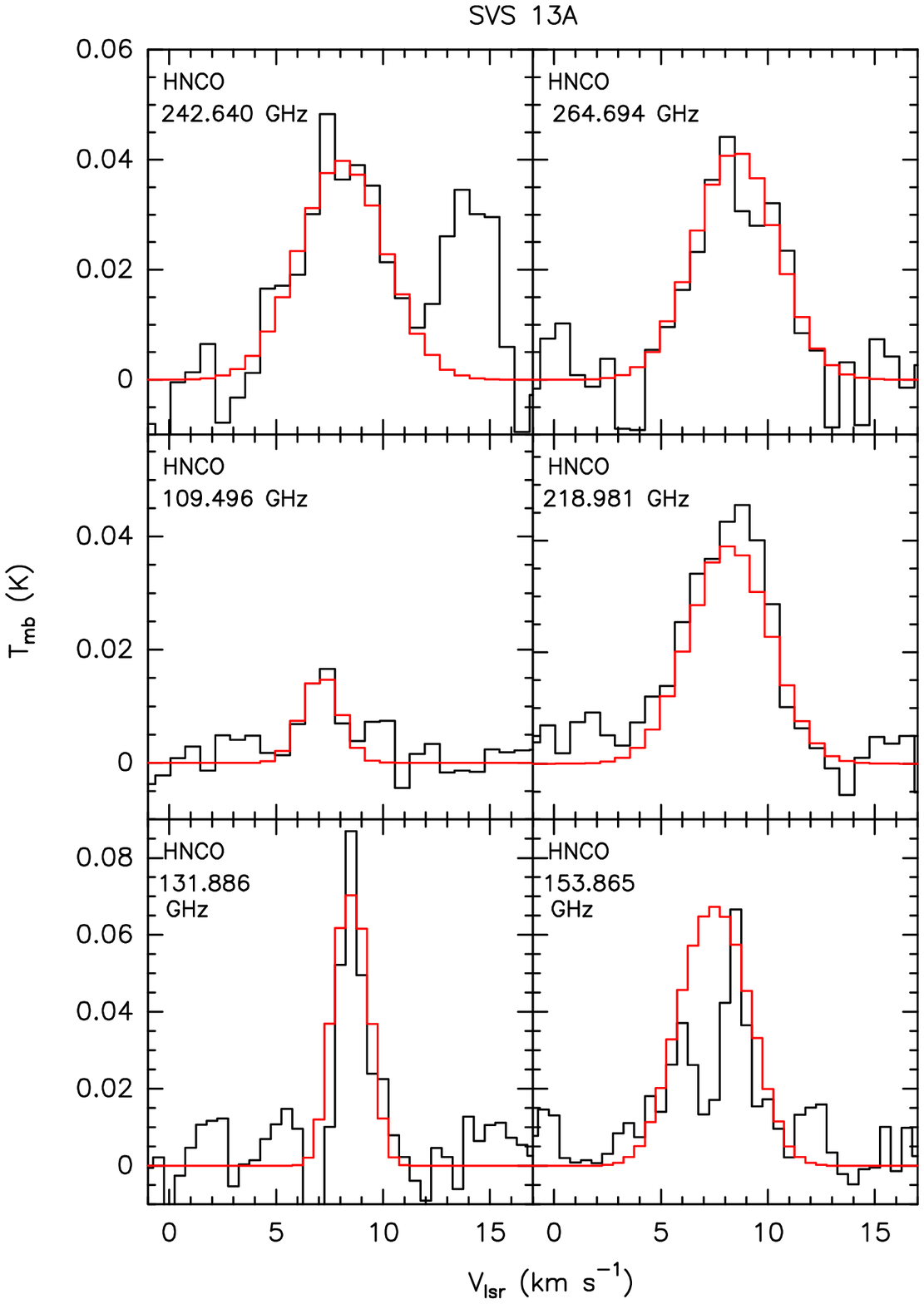} & \includegraphics[scale=0.55]{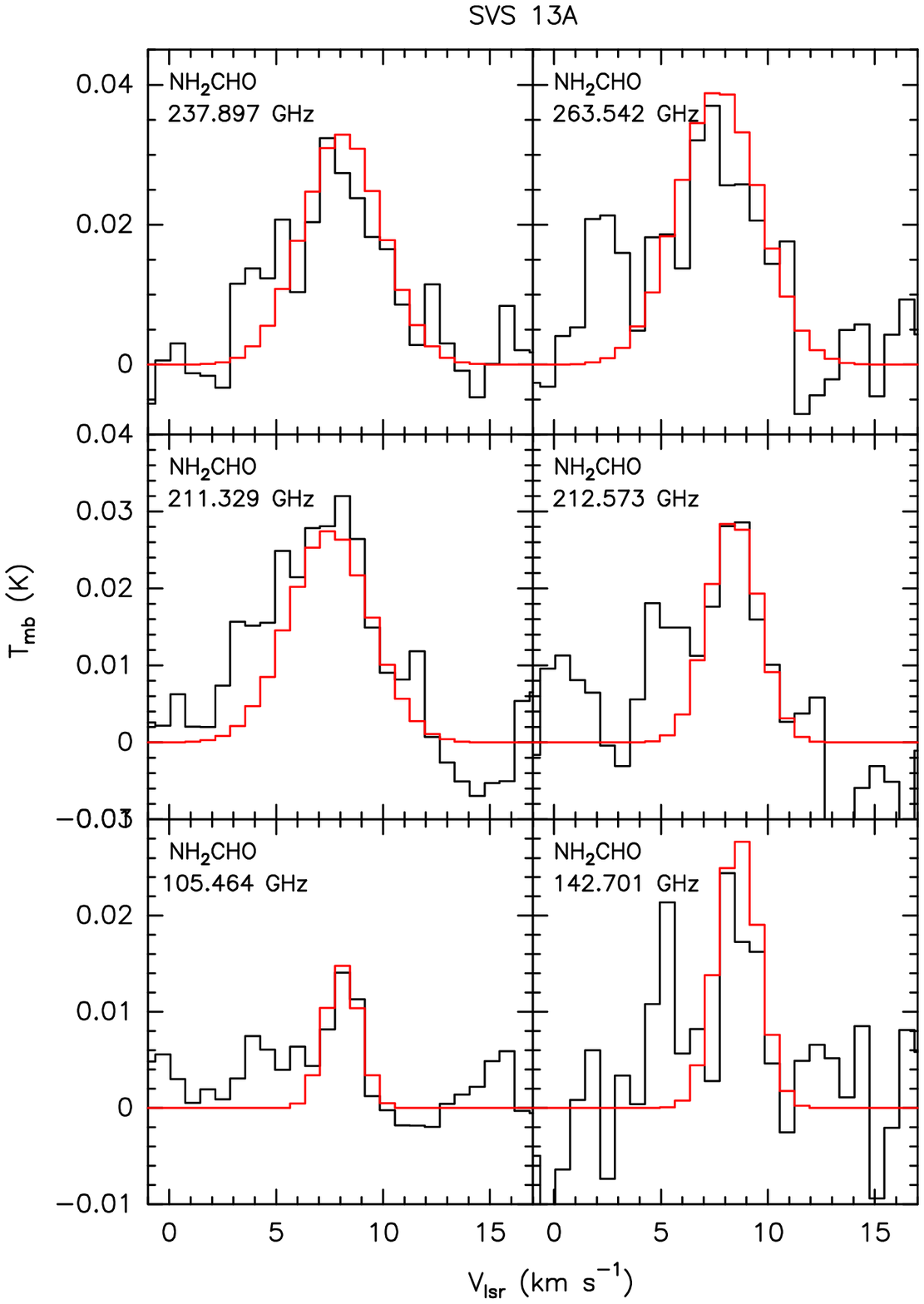}\\
\end{tabular}
  \caption{Sample of HNCO (\textit{left}) and NH$_2$CHO (\textit{right}) observed spectral lines (black) in IRAS~4A and SVS13A (compact solution), and the spectra predicted by best fit LTE model (red).}
  \label{fspt2}
\end{figure*}

\clearpage

\begin{figure*}
\centering
\begin{tabular}{lr}
   \includegraphics[scale=0.55]{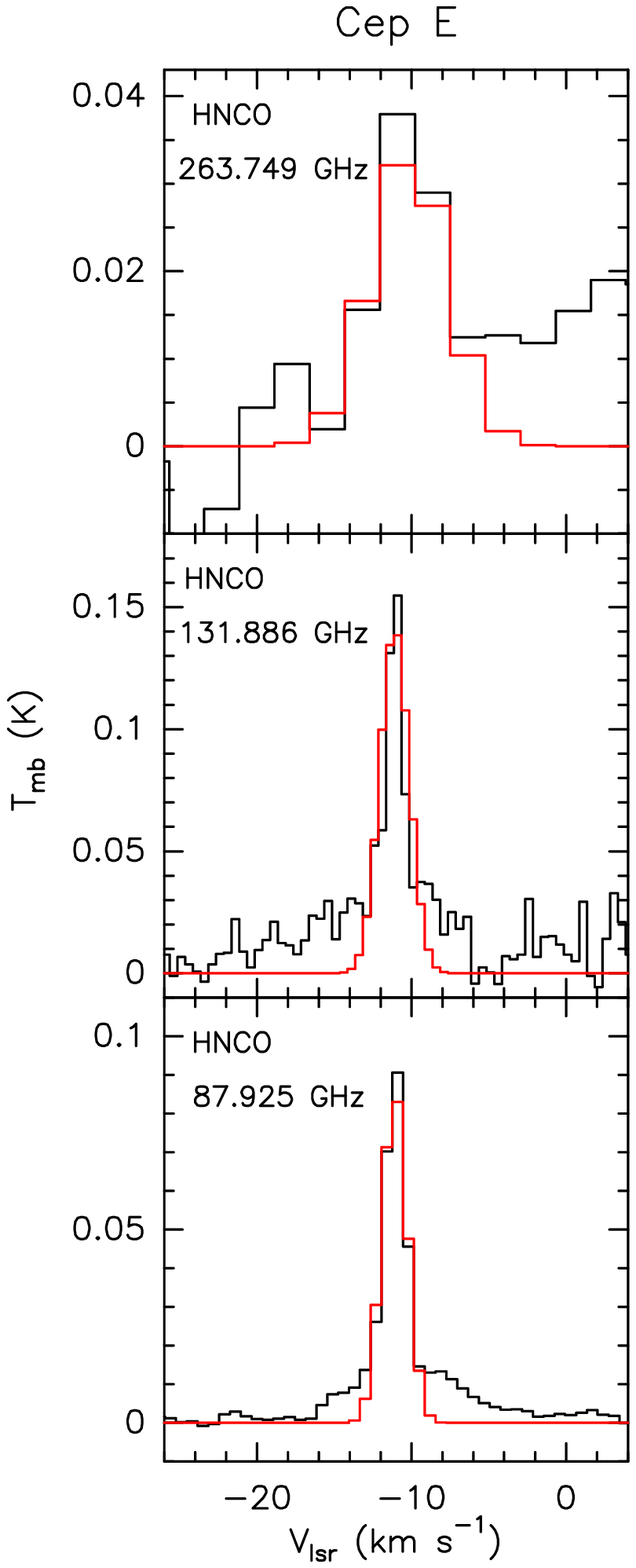} & \includegraphics[scale=0.55]{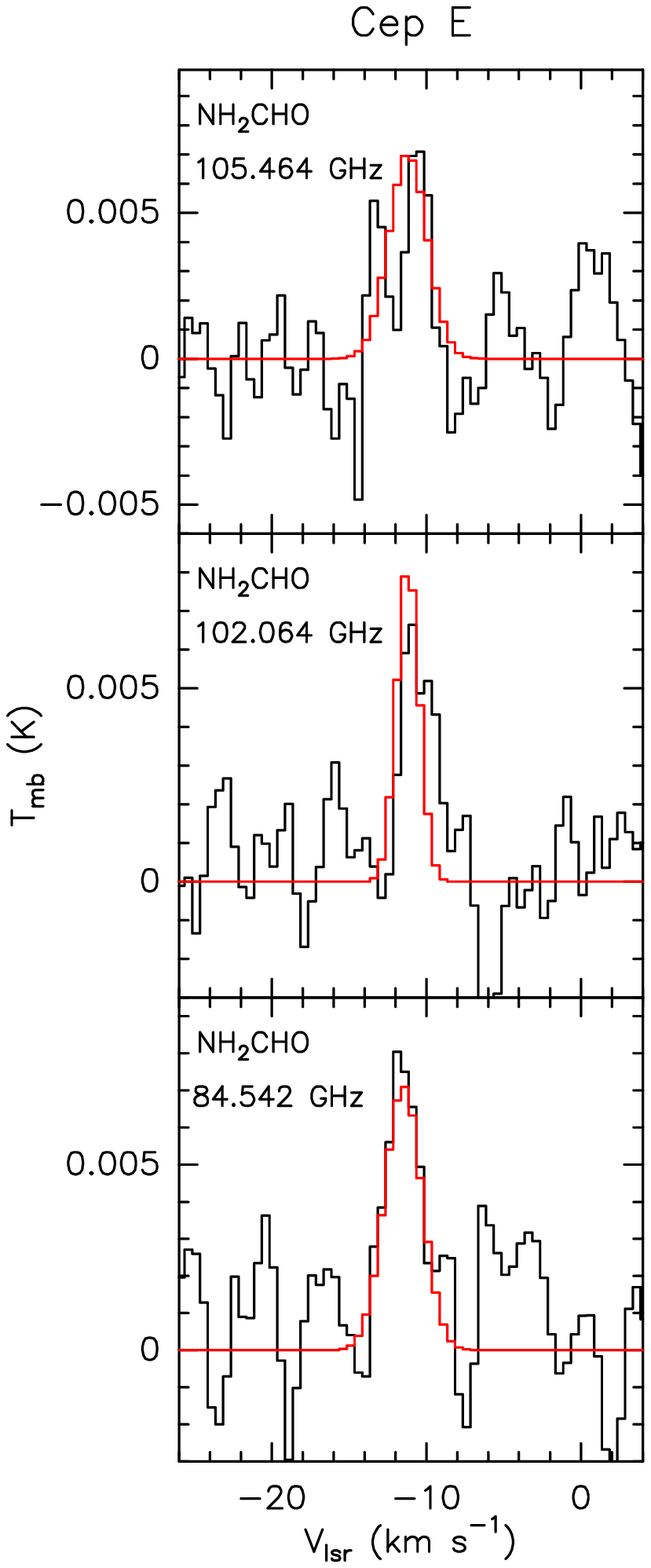}\\
   \includegraphics[scale=0.55]{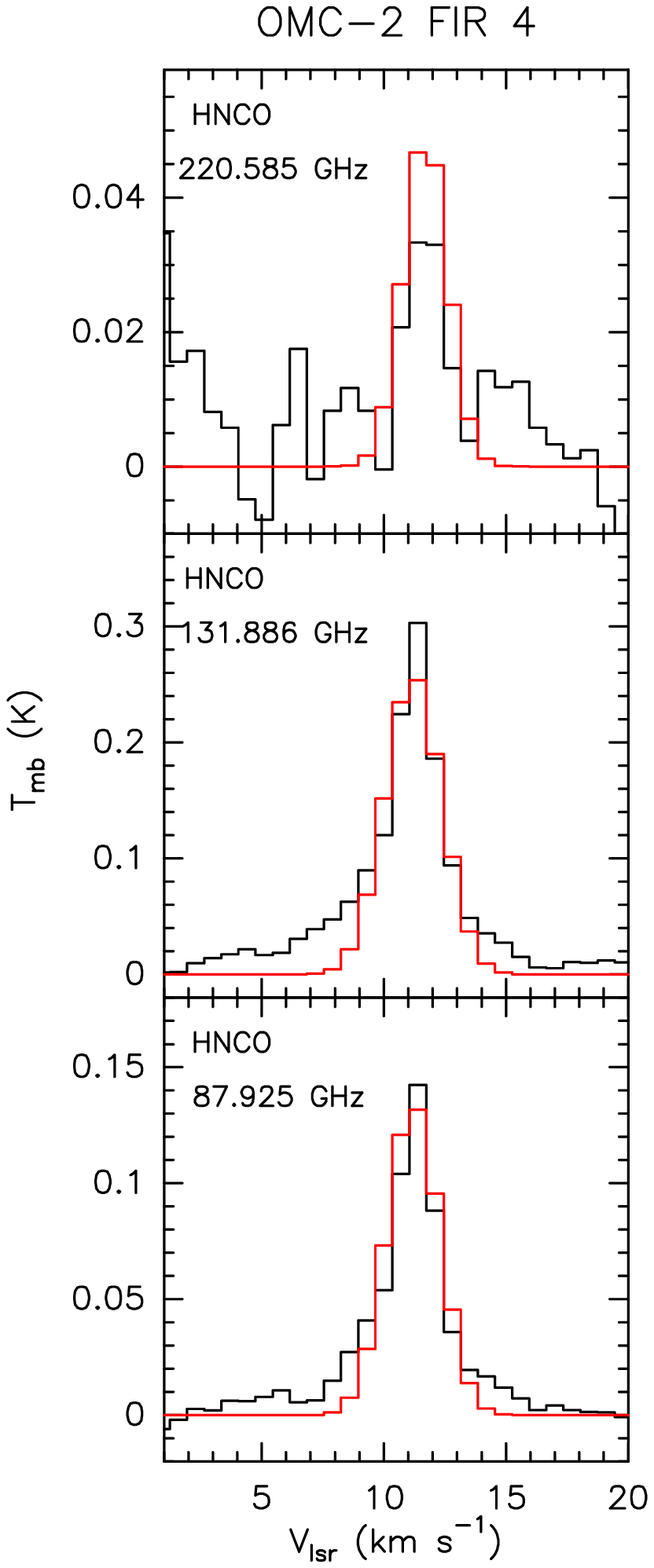} & \includegraphics[scale=0.55]{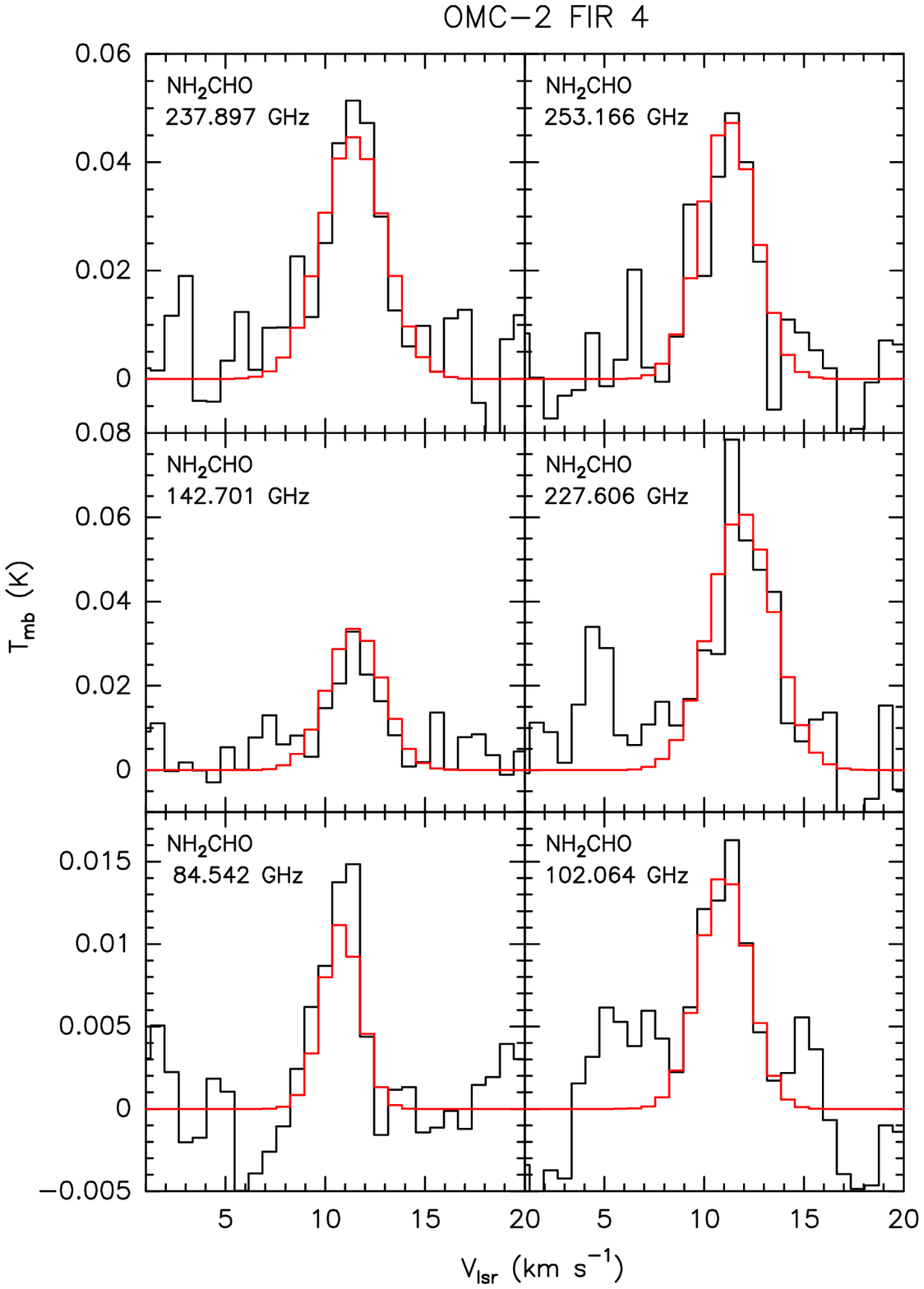}\\
\end{tabular}
  \caption{Sample of HNCO (\textit{left}) and NH$_2$CHO (\textit{right}) observed spectral lines (black) in Cep~E and OMC-2~FIR~4 (extended solutions), and the spectra predicted by best fit LTE model (red).}
  \label{fspt3}
\end{figure*}

\begin{figure*}
\centering
\begin{tabular}{lcr}
  \includegraphics[scale=0.69]{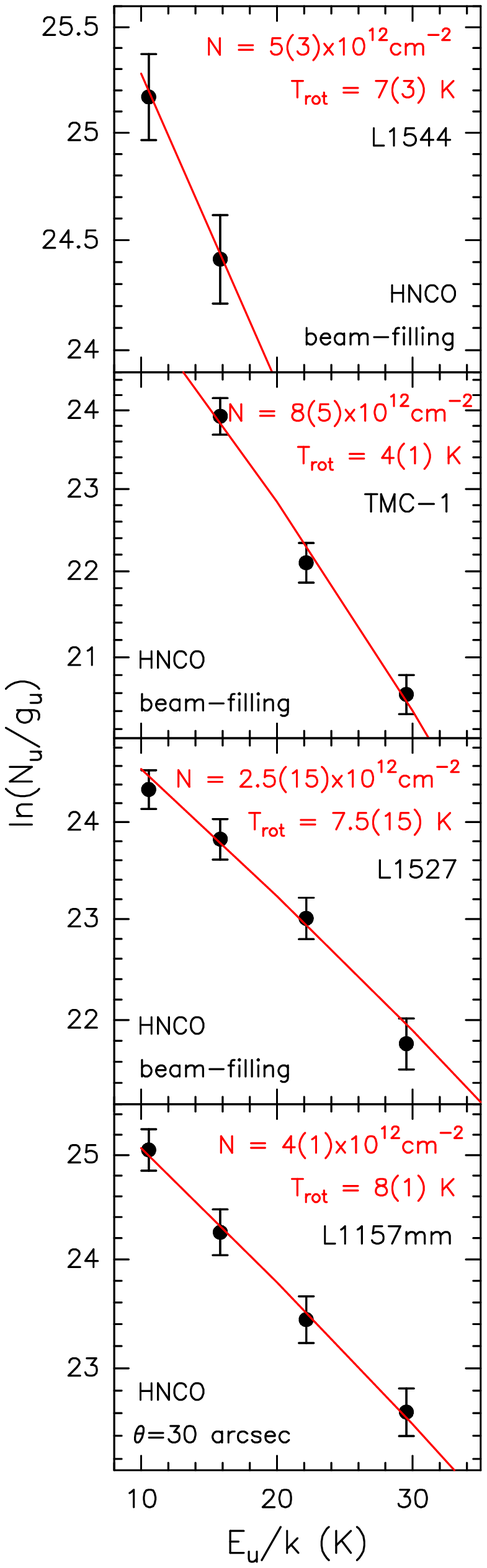} & \includegraphics[scale=0.69]{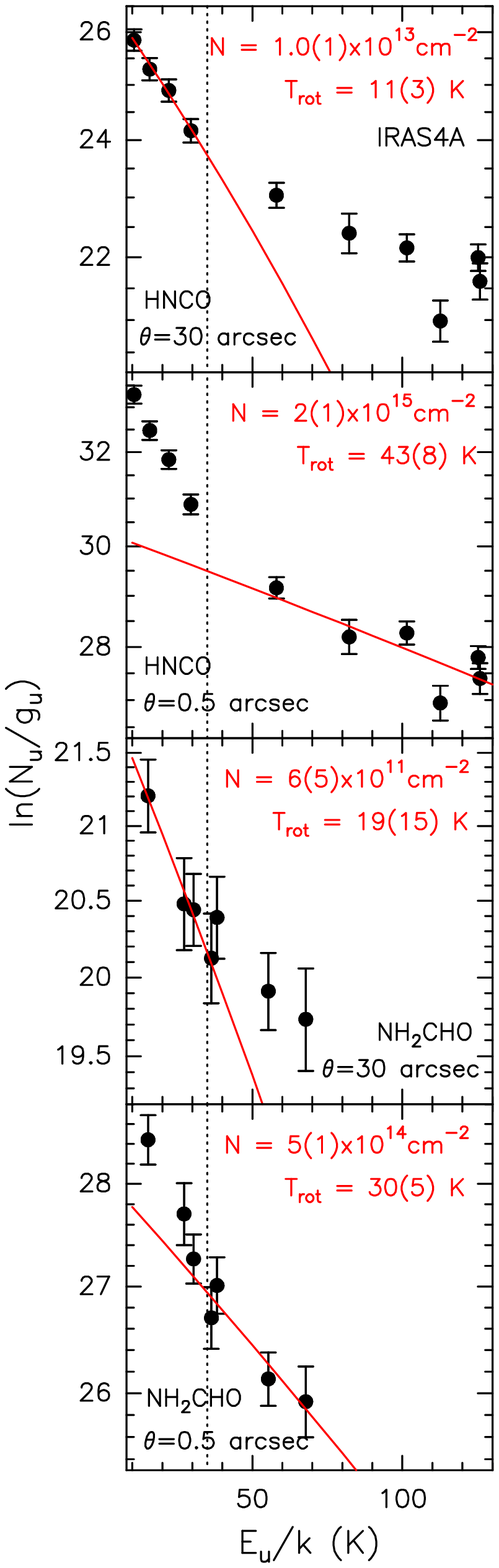} & \includegraphics[scale=0.69]{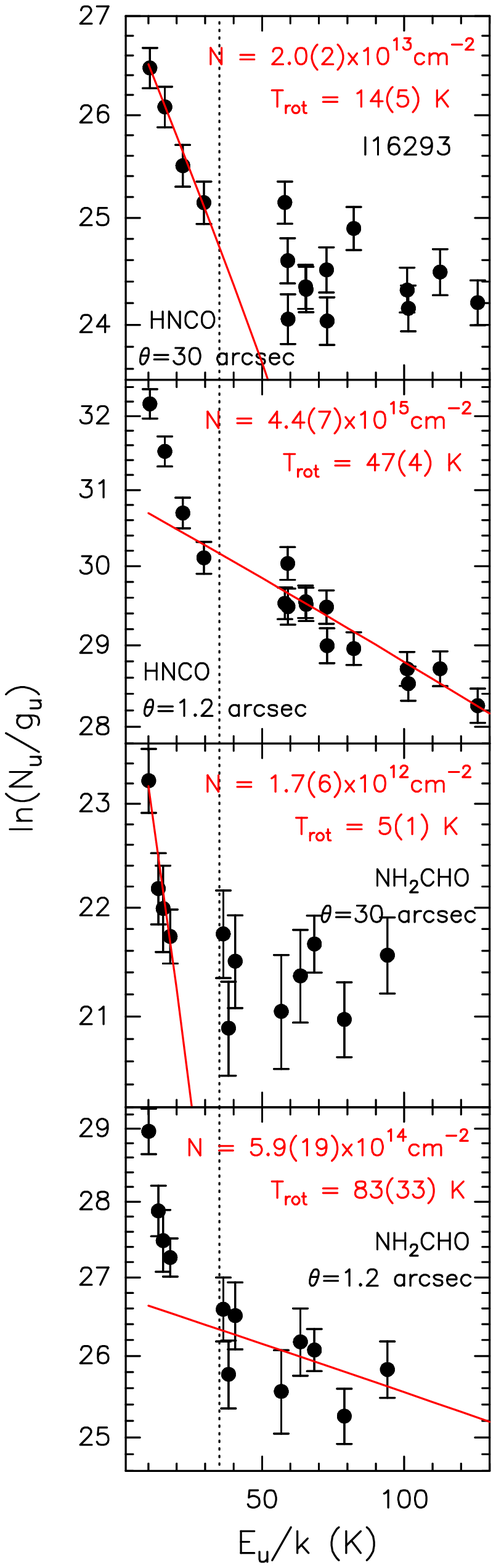}\\
\end{tabular}
  \caption{Rotational diagrams for L1544, TMC-1, L1527, and L1157mm (\textit{left}), IRAS~4A (\textit{middle}), and I16293 (\textit{right}). Data points are depicted in black. The red lines correspond to the best fit to the data points. The dashed vertical lines in the middle and right panels indicate the upper-level energy (35~K) at which the division of the 2-component fitting was made.}
  \label{frot1}
\end{figure*}

\begin{figure*}
\centering
\begin{tabular}{lcr}
  \includegraphics[scale=0.69]{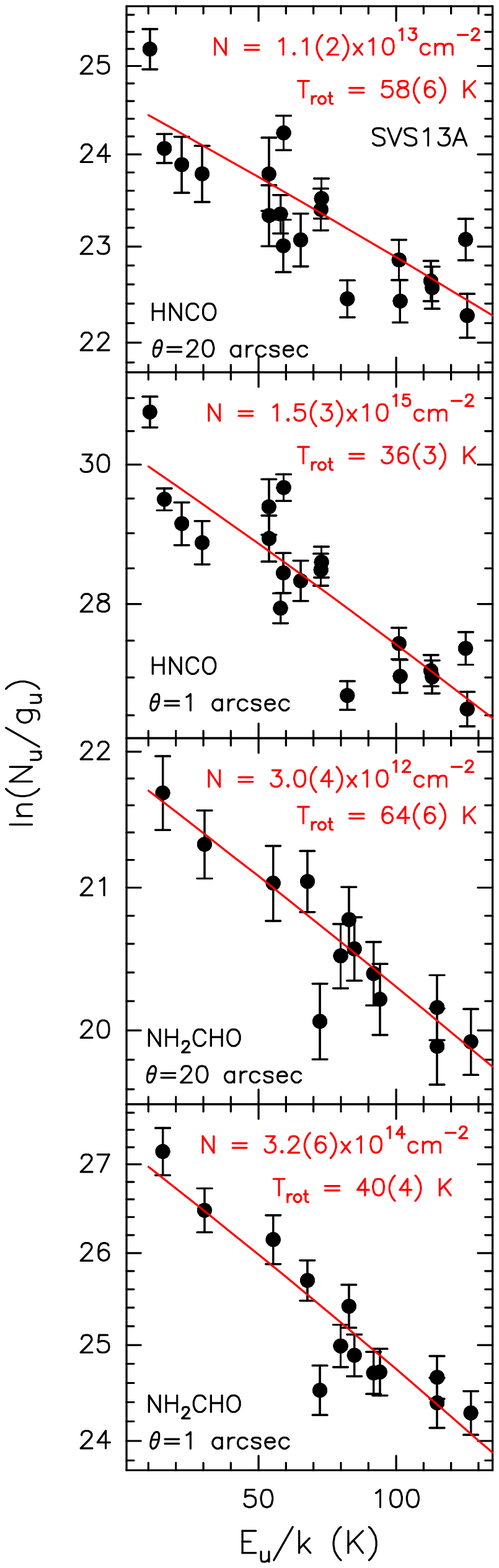} & \includegraphics[scale=0.69]{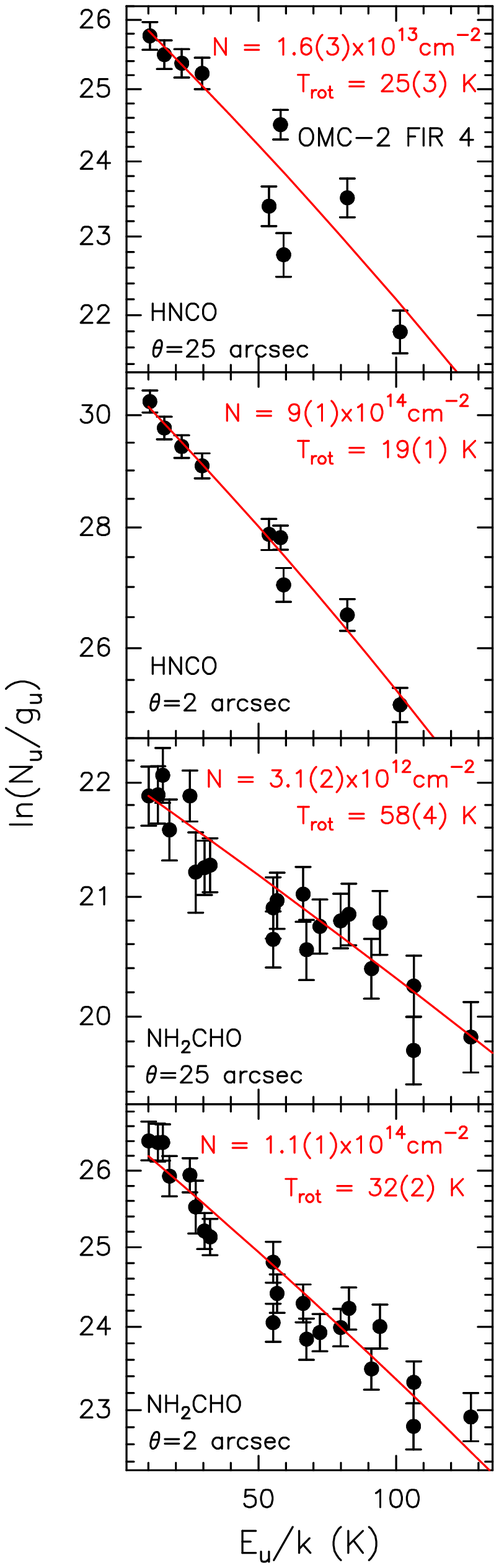} & \includegraphics[scale=0.69]{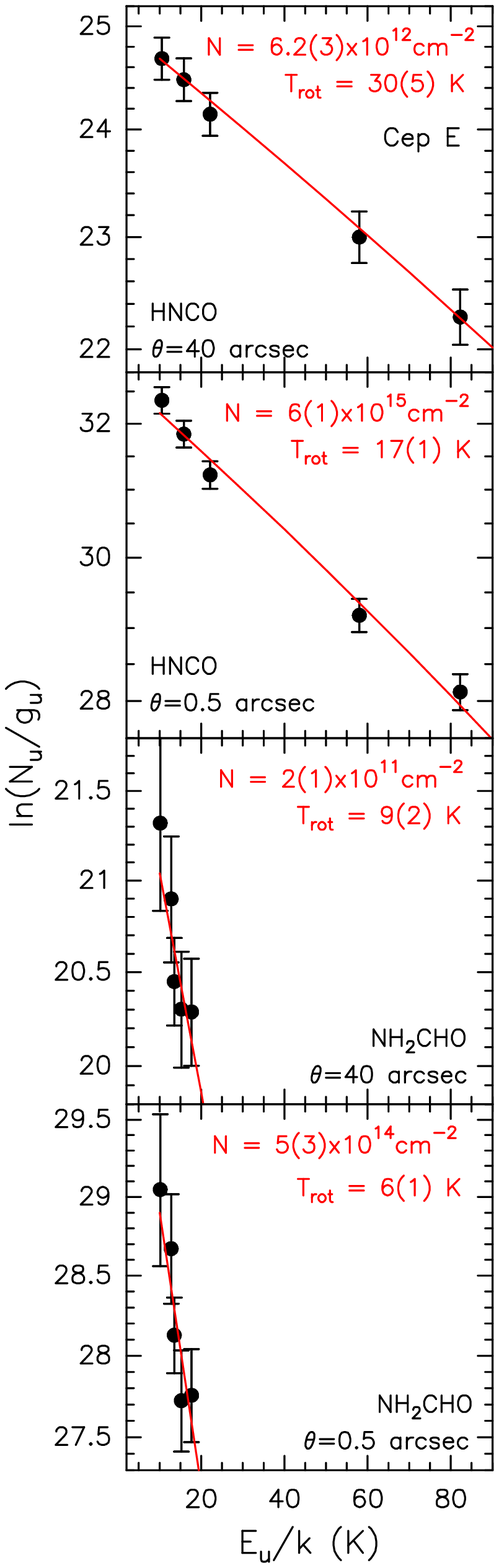}\\
\end{tabular}
  \caption{Rotational diagrams for SVS13A (\textit{left}), OMC-2~FIR~4 (\textit{middle}), and Cep~E (\textit{right}). Data points are depicted in black. The red lines correspond to the best fit to the data points. .}
  \label{frot2}
\end{figure*}

\end{document}